%% file: hexabsIclean.tex
\documentclass{llncs} 

\usepackage{amssymb}
\usepackage{algorithm}
\usepackage{color}
\usepackage{graphicx}
\usepackage{mathptmx} 

\newcommand{\bs}{\backslash}
\newcommand{\PF}{{\bf Proof: }}
\newcommand{\QED}{\hspace*{\fill}{$\Box$}\medskip}

\def\notat#1{{$#1$}\marginpar{\raggedright{\small $#1$}}}
\def\term#1{{\em #1}\marginpar{\raggedright{\small\it #1}}}

\newcommand{\BW}{\mathcal{B}}
\newcommand{\AS}{\mathcal{A}}

\newcommand{\convex}{\mbox{\sc Right}}
\newcommand{\concave}{\mbox{\sc Left}}

\newcommand{\Bo}{B'}
\newcommand{\Bt}{B''}

\newcommand{\EC}{\mathcal{C}}
\newcommand{\npa}{p}
\newcommand{\elcyc}{elementary cycle}

\newcommand{\GPA}{G^{W}}

\newtheorem{thm}{Theorem}
\newtheorem{lem}[thm]{Lemma}

\newtheorem{corol}[thm]{Corollary}
\newtheorem{propo}[thm]{Proposition}

\begin{document}

\title{Counting hexagonal patches and independent sets in circle graphs}
\titlerunning{counting hexagonal patches}
\date{\today}

\author{Paul Bonsma\inst{1}\thanks{Both authors are supported by the Graduate School ``Methods for Discrete Structures'' in Berlin, DFG grant GRK 1408.} \and Felix Breuer\inst{2}}

\institute{
Technische Universit\"{a}t Berlin, Institut f\"{u}r Mathematik,\\
Sekr. MA 5-1, Stra\ss{}e des 17. Juni 136, 10623 Berlin, Germany.\\ \texttt{bonsma@math.tu-berlin.de}
\and
Freie Universit\"{a}t Berlin, Institut f\"{u}r Mathematik,\\
Arnimallee 3, 14195 Berlin, Germany.\\
\texttt{felix.breuer@fu-berlin.de}
}

\maketitle

\begin{abstract}
A hexagonal patch is a plane graph in which inner faces have length 6, inner vertices have degree 3, and boundary vertices have degree 2 or 3. 
We consider the following counting problem: given a sequence of twos and threes, how many hexagonal patches exist with this degree sequence along the outer face?
This problem is motivated by the enumeration of benzenoid hydrocarbons and fullerenes in computational chemistry.
We give the first polynomial time algorithm for this problem. We show that it can be reduced to counting maximum independent sets in circle graphs, and give a simple and fast algorithm for this problem. 
\end{abstract}

\section{Introduction}

The notions used and problems introduced in this section are defined more formally in Section~\ref{sec:prelim}.
A plane graph $G$ is a graph together with a fixed planar embedding in the plane. The unbounded face is called the {\em outer face} and the other faces are called {\em inner faces}. The boundary of the outer face is simply called the {\em boundary} of $G$.
A {\em hexagonal patch} is a 2-connected plane graph in which all inner faces have length 6, boundary vertices have degree 2 or 3, and non-boundary vertices have degree 3. These graphs are also known as {\em fusenes}~\cite{GHZ02}, {\em hexagonal systems}~\cite{DFG01}, {\em polyhexes}~\cite{Gra03} and {\em $(6,3)$-polycycles}~\cite{DDS08} in the literature. These graphs are well-studied in mathematical and computational chemistry since they model benzenoid hydrocarbons and graphite fragments (see e.g.~\cite{GHZ02} and the references therein). A central question is that of enumerating hexagonal patches, either of a given size~\cite{BCH03}, or with a given boundary form.

\begin{figure}
\centering
\scalebox{0.8}{$\input{nontrivinstance.pstex_t}$}
\caption{Two different patches with the same boundary code.}
\label{fig:nontriv}
\end{figure}
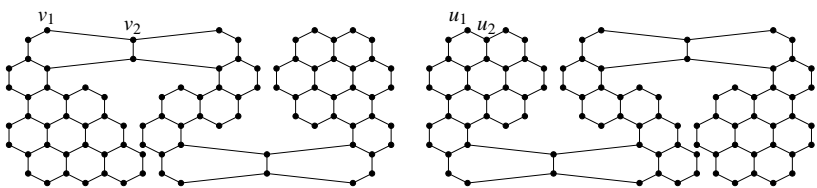

A sequence $x_0,\ldots,x_{k-1}$ of twos and threes is a {\em boundary code} of a hexagonal patch $G$ if there is a way to label the boundary vertices of $G$ with $v_0,\ldots,v_{k-1}$ such that $v_0,\ldots,v_{k-1},v_0$ is a boundary cycle of $G$, and the degree $d(v_i)=x_i$ for all $i$. Note that cyclic permutations and/or inversions of the sequence can yield different boundary codes for the same patch, but for the question we study this fact is not important.
It is well-known and easily observed using Euler's formula that the boundary code of a hexagonal patch satifies $d_2-d_3=6$, where $d_i$ is the number of boundary vertices with degree $i$. We define the parameters $d_2(X)$ and $d_3(X)$ also for sequences $X$ of twos and threes, as expected.
One may ask whether a hexagonal patch exists that satisfies a given boundary code.
A result by Guo, Hansen and Zheng~\cite{GHZ02} shows that this question is not as easy as was first expected: in Figure~\ref{fig:nontriv} their example is shown which shows that 
different patches may exist with the same boundary code. This can be verified by comparing the degree of $v_1$ with $u_1$, $v_2$ with $u_2$, etc. 
Our drawing of this graph is taken from~\cite{BDN05}. (In~\cite{GHZ02} it is also shown that although multiple solutions may exist, they all have the same size.)
Therefore the following question should be asked: given a sequence $S$ of twos and threes, {\em how many} hexagonal patches exist with boundary code $S$? This counting problem is called {\em Hexagonal Patch}. Guo et al~\cite{GHZ02} and Graver~\cite{Gra03} give conditions for when solutions are unique, if they exist. Deza et al~\cite{DFG01} give an algorithm for deciding whether at least one solution exist. The complexity of their algorithm is however superexponential.
In addition they give a polynomial time algorithm for a very restricted case (see Section~\ref{sec:representations}).
These results have been generalized to various problem variants, mainly by varying the conditions on the face lengths and vertex degrees, see e.g.~\cite{DDS08,BC09,BDN05,BGJ09}. However, the question whether the counting problem can be answered efficiently remained open.

\medskip

{\em In this paper we show that the counting problem Hexagonal Patch can be solved in time $O(k^3)$ where $k$ is the length of the sequence.} This is surprising since the number of solutions may be exponential in $k$, as can easily be seen by generalizing the example from Figure~\ref{fig:nontriv}.
Therefore, we can only return the {\em number} of solutions in polynomial time, and not return a {\em list} of all corresponding patches. The algorithm can however be extended to generate all patches in time $k^{O(1)}\cdot O(n)$, where $n$ is the number of returned solutions.
We remark that it is not hard to generalize our result to the generalizations introduced in~\cite{BDN05}:
A 2-connected plane graph is an {\em $(m,k)$-patch} if all inner faces have length $k$, inner vertices have degree $m$ and boundary vertices have degree at most $m$. 
Our methods work for instance for $(4,4)$ and $(6,3)$-patches in addition to $(3,6)$-patches, but for simplicity we restrict to hexagonal patches.

An additional motivation for this result is the following: {\em fullerene patches} generalize hexagonal patches by also allowing 5-faces in addition to 6-faces. Such patches model fragments of fullerene molecules, and therefore their enumeration is another important problem from computational chemistry. Fullerene molecules have at most twelve 5-faces. 
The current result is an essential ingredient for the result we give in a second paper~\cite{BB2}, where we give a polynomial time algorithm for deciding whether a given boundary code belongs to a fullerene patch with at most five 5-faces.

Our algorithm is based on the following idea: with a few intermediate steps, we transform the problem Hexagonal Patch to the problem of counting maximum independent sets in circle graphs. A \term{circle graph} $G$ is the intersection graph of chords of a circle (detailed definitions are given below). Algorithms are known for the optimization problem of finding maximum independent sets in circle graphs~\cite{Gav73}, but counting problems on circle graphs have not been studied to our knowledge.

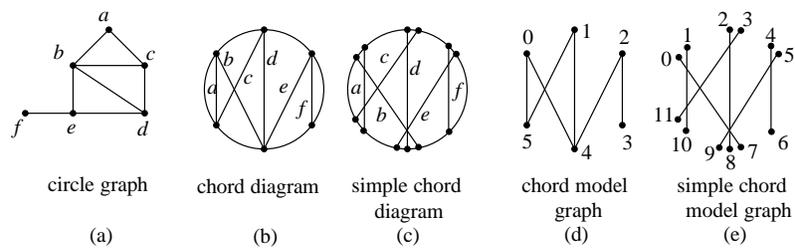
\begin{figure}
\centering
\scalebox{1}{$\input{circlegraph.pstex_t}$}
\caption{A circle graph, (simple) chord diagram and (simple) chord model graph.}
\label{fig:circlegraph}
\end{figure}

{\em In this paper we give a simple dynamic programming algorithm for counting independent sets in circle graphs}. In addition this algorithm improves the complexity for the optimization problem.
Circle graphs can be represented as follows
(see Figure~\ref{fig:circlegraph}(a),(b)): Every vertex of $G$ is associated with a {\em chord} of a circle drawn in the plane, which is a straight line segment between two points on the circle, such that two vertices are adjacent if and only if the two chords overlap (possibly only in a common end).
We will represent chord diagrams with graphs as follows (see Figure~\ref{fig:circlegraph}(d)).
Number the points on the circle that are ends of chords with $0,\ldots,k-1$, in order around the circle, and view these as vertices. View a chord from $i$ to $j$ as an edge $ij$. 
Call the resulting graph $G'$ the \term{chord model graph}.
Note that (maximum) independent sets of the circle graph correspond bijectively to \term{(maximum) planar matchings} or (M)PMs of $G'$, which are (maximum) matchings $M$ that do not contain edges $\{i,j\}$ and $\{x,y\}$ with $i<x<j<y$.
Hence counting MPMs in $G'$ is polynomially equivalent to counting maximum independent sets in circle graphs.

Circle graphs are extensively studied and generalize permutation graphs and distance hereditary graphs, see e.g.~\cite{BLS99}. Recognizing them and constructing a chord representation can be done in polynomial time~\cite{Bou87,GSH89}, and the current fastest algorithm uses time $O(n^2)$, where $n$ is the number of vertices~\cite{Spi94}. A number of problems that are NP-hard on general graphs are easy on circle graphs, such as in particular finding maximum independent sets~\cite{Gav73,Sup87,Val03,NLG09}.

The first algorithm for the optimization problem by Gavril~\cite{Gav73} has time complexity $O(m^3)$, where $m$ is the number of vertices of the circle graph, which is the number of edges of the corresponding chord model graph.
This was improved to $O(m^2)$ by Supowit~\cite{Sup87}. Recently this has been improved further by Valiente~\cite{Val03} in the way we will explain now. 
All of these algorithms work with the chord model graph (or chord diagram), and as a first step, transform it into a 1-regular graph as shown in Figure~\ref{fig:circlegraph}(e):
for a vertex of degree $d$, $d$ new vertices are introduced, and the $d$ incident edges are distributed among these in such a way only one of these edges can appear in a PM of $G$. This does not change the size and number of MPMs. The resulting graph $G$ has $2m$ vertices and $m$ edges, and is called the {\em simple chord model graph}. We assume the vertices are numbered $0,\ldots,2m-1$, in the proper order. Then the {\em length} of an edge $ij\in E(G)$ is $|j-i|$. The algorithm from~\cite{Val03} has complexity $O(l)$, where $l$ is the sum of all edge lengths of the simple chord model graph obtained this way. Clearly this is at most $O(m^2)$, and in many cases better. However, when {\em dense} chord model graphs are given on $n$ vertices and $m\in \Omega(n^2)$ edges, this algorithm may need $\Omega(n^4)$ steps. 
Our transformation from Hexagonal Patch yields a chord model graph $G'$, which in fact may be dense.

We give a simple algorithm with complexity $O(nm)$, which not only determines the size of a MPM, but also counts the number of MPMs of the chord model graph. 
This improvement in time complexity is possible by working with arbitrary degrees, and not using the simple chord model graph, in contrast to all previous algorithms for this problem~\cite{Val03}.

The outline of the paper is as follows. In Section~\ref{sec:prelim} we give definitions, and a precise formulation of the problem. In Section~\ref{sec:representations} we define locally injective homomorphisms to the hexagonal lattice (the brickwall) as a way of representing problem instances and solutions and reduce the counting problem to a problem on walks in the brickwall. 
In Section~\ref{sec:assignments} we reduce that problem to that of counting {\em proper assignment sets} of the walk, which is in fact the problem of counting MPMs in chord model graphs.
In Section~\ref{sec:circlegraphs} we present our algorithm for counting MPMs, and in Section~\ref{sec:alg} we give a summary of our algorithm for Hexagonal Patch.
We end in Section~\ref{sec:discussion} with a discussion, where we also discuss a similar problem from topology.
Statements for which proofs are omitted are marked with a star, the proofs appear in the appendix.

\section{Preliminaries}
\label{sec:prelim}

For basic graph theoretic notions not defined here we refer to~\cite{Die}.
A \term{walk} of {\em length $k$} in a (simple) graph $G$ is a sequence of $k+1$ vertices $v_0,\ldots,v_k$ such that $v_i$ and $v_{i+1}$ are adjacent in $G$ for all $i\in \{0,\ldots,k-1\}$.
$v_1,\ldots,v_{k-1}$ are the \term{internal} vertices and $v_0,v_k$ the \term{end} vertices of the walk. The walk is \term{closed} if $v_0=v_k$.
Throughout this paper we will in addition assume that $v_{i-1}\not=v_{i+1}$ for all $i\in\{1,\ldots,k-1\}$, and if the walk is closed, $v_1\not=v_{k-1}$ (i.e. we will assume walks {\em do not turn back}).
If $v_i\not=v_j$ for all $i\not=j$ 
then the walk is a \term{path}. If the walk is closed and $v_i\not=v_j$ for all distinct $i,j\in\{0,\ldots,k-1\}$ then it is also called a \term{cycle}. A cycle of length $k$ is also called a \term{$k$-cycle}.
For a walk $W=v_0,\ldots,v_k$, \notat{W_x} denotes $v_x$.
If $W$ is a closed walk, then $W_x$ denotes $v_{x \bmod k}$. 
We will also talk about the {\em vertices} and {\em edges} of a walk, which are defined as expected. In a slight abuse of terminology, the graph consisting of these vertices and edges will also be called a walk (or path or cycle if applicable).

Let $H$ be a hexagonal patch, and $B$ be a boundary cycle of $H$ of length $k$. Let $X=x_0,\ldots,x_{k-1}$ be a sequence of twos and threes. We say that the tuple $(H,B)$ is a {\em solution for the boundary code $X$} if $d(B_i)=x_i$ for all $i\in\{0,\ldots,k-1\}$.
Two solutions $(H,B)$ and $(H',B')$ are considered \term{equivalent} if there is an isomorphism $\psi$ from $H$ to $H'$ such that $\psi(B_i)=B'_i$ for all $i$.
Formally, when we ask for the number of {\em different} pairs $(H,B)$ that satisfy some property, we want to know how many equivalence classes contain a pair $(H,B)$ satisfying this property.
The counting problem Hexagonal Patch is now defined as follows: given a sequence $X$, how many different solutions $(H,B)$ to $X$ exist?

\section{From Boundary Codes to Walks in the Brickwall}
\label{sec:representations}

\begin{figure}
\centering
\scalebox{0.40}{$\input{brickwall_nolabels.pstex_t}$}
\caption{The brickwall $\BW$.}
\label{fig:brickwalls}
\end{figure}
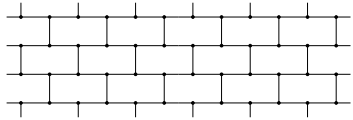

An (infinite) 3-regular plane graph where every face has length 6 is called a \term{brickwall}. It can be shown that the facial cycles are the only 6-cycles of a brickwall, and that all brickwalls are isomorphic.

We will use \notat{\BW} to denote the brickwall as drawn in Figure~\ref{fig:brickwalls}. Edges that are horizontal (vertical) in this drawing are called the \term{horizontal} (\term{vertical}) edges of $\BW$.
Paths consisting of horizontal edges are called \term{horizontal paths}. Two vertices joined by a horizontal path are said to have the same \term{height}.

The reason that we study brickwalls is because the following mapping of hexagonal patches into them is very useful.
Let $H$ be a hexagonal patch. A \term{locally injective homomophism (LIH)} of $H$ into $\BW$ is a mapping of the vertices of $H$ to vertices of $\BW$, such that adjacent vertices are mapped to adjacent vertices in $\BW$, and such that all neighbors of any vertex in $H$ are mapped to different vertices in $\BW$.
Since the shortest cycles in $\BW$ are of length 6, a LIH into $\BW$ maps 6-cycles to 6-cycles. Since the faces of $\BW$ are the only 6-cycles in $\BW$, we see that a LIH of $H$ into $\BW$ also maps inner faces to faces.

Loosely speaking, the idea behind these mappings is as follows. 
Let $H$ be a hexagonal patch of which we fix a boundary cycle $B$. When we map $H$ with a LIH $\phi$ into $\BW$, then the boundary $B$ is mapped to some walk $W$ in $\BW$. 
But now it can be shown that this walk $W$ is only determined by the choice of the initial vertices and the boundary code of $H$.
Hence instead of asking how many hexagonal patches exist with a certain boundary code, we may ask how many patches exist that can be mapped properly to the brickwall, such that the boundary coincides with the walk that is deduced from the boundary code. 
Below we will go into more detail.

The technique of mapping patches to brickwalls is not new, and is actually considered folklore to some extent~\cite{DFG01}. 
For instance, Deza et al~\cite{DFG01} observe that Hexagonal Patch can be solved in polynomial time if the LIH is bijective, and Graver~\cite{Gra03} shows that the problem Hexagonal Patch can only have multiple solutions if there is a brickwall vertex that has at least three preimages in such a LIH.
We will however study these mappings more in more detail than has been done before, and develop new concepts, and prove new statements which we feel are of independent interest.

Let $W$ be a walk in a 3-regular plane graph $G$. We say \term{$W$ makes a right (left) turn at $i$} when 
edge $W_iW_{i-1}$ immediately follows edge $W_iW_{i+1}$ in the clockwise (anticlockwise) order around $W_i$.
Note that since we assume that walks do not turn back and $G$ is 3-regular, $W$ makes either a left or a right turn at every $i$.

{\em Walk construction:}
Using a given sequence $x_0,\ldots,x_{k-1}$ of twos and threes, we construct a walk $W=v_0,\ldots,v_k$ in $\BW$ as follows. 
For $v_0$ and $v_1$, choose two (arbitrary) adjacent vertices.
For $i\geq 1$, choose $v_{i+1}$ such that $W$ makes a left turn at $i$ if $x_i=3$, and makes a right turn at $i$ if $x_i=2$.

Let $W$ be a closed walk in $\BW$ of length $k$, $H$ be a hexagonal patch, $\phi$ a LIH from $H$ to $\BW$ and $B$ a boundary walk of $H$ of length $k$. Then the tuple $(H,\phi,B)$ is said to be a \term{solution for $W$} when   $\phi(B_i)=W_i$ for all $i$.
Two solutions $S=(H,\phi,B)$ and $S'=(H',\phi',B')$ are considered to be \term{equivalent} if and only if there is an isomorphism $\psi$ from $H$ to $H'$ such that $\psi(B_i)=B'_i$ for all $i$. We say that $\psi$ is an (or demonstrates the) equivalence between $S$ and $S'$.
The LIH $\phi$ allows us to use the terminology defined for $\BW$ for the graph $H$ as well; we will for instance call edges of $H$ {\em horizontal} or {\em vertical} if their images under $\phi$ are horizontal or vertical, respectively. 

Let the boundary $B$ of a hexagonal patch $H$ be mapped to the closed walk $W$ in $\BW$ by the LIH $\phi$. This is a \term{clockwise solution} if and only if for every $i$, 
  $d(B_i)=2$  if $W$ makes a right turn at $i$, and
  $d(B_i)=3$  if $W$ makes a left turn at $i$.
It is {\em anticlockwise} when these conditions are reversed.
Let \notat{\convex(W)} and \notat{\concave(W)} denote the number of indices $i\in \{0,\ldots,k-1\}$ such that $W$ makes a right turn or left turn at $i$, respectively. 
The \term{turning number} of $W$ is \notat{t(W)}$=(\convex(W)-\concave(W))/6$.
Using the fact that for a solution $(H,\phi,B)$, $\phi$ maps faces of $H$ to faces of $\BW$, it can be shown that every solution is either clockwise or anticlockwise. Since a hexagonal patch has $d_2-d_3=6$ ($d_i$ is the number of degree $i$ vertices on the boundary), Lemma~\ref{lem:cw_or_acw} then follows. Variants of Lemma~\ref{lem:unique_phi} have been proved in~\cite{BDN05,Gra03}.

\begin{lem}[*]
\label{lem:cw_or_acw}
Let $W$ be a closed walk in $\BW$.
If $t(W)=1$, then every solution to $W$ is clockwise. If $t(W)=-1$ then every solution to $W$ is anticlockwise. If $t(W)\not\in \{-1,1\}$, then no solution exists.
\end{lem}

\begin{lem}
\label{lem:unique_phi}
Let $(H,B)$ be a solution to a boundary code $X$ and let $W$ be a walk in $\BW$ that is constructed using $X$. Then there exists a unique LIH $\phi$ such that $(H,\phi,B)$ is a clockwise solution to $W$.
\end{lem}

Because of Lemma~\ref{lem:unique_phi}, we may rephrase the problem Hexagonal Patch in terms of solutions $(H,\phi,B)$ to a closed walk $W$ in the brickwall. 

\begin{thm}
\label{thm:equiv_boundary_walk}
The number of different (hexagonal) solutions for a boundary code $X$ with $d_2(X)-d_3(X)=6$ is the same as the number of different clockwise solutions for the walk $W$ in $\BW$ that is constructed using $X$.
\end{thm}
\PF
For any solution $(H,B)$ for $X$, a unique LIH $\phi$ exists such that $(H,\phi,B)$ is a clockwise solution to $W$ (Lemma~\ref{lem:unique_phi}). For any clockwise solution $(H,\phi,B)$ to $W$, the characterization of clockwise solutions and the construction of $W$ shows that $(H,B)$ is a solution to $X$. 
(Since $d_2(X)-d_3(X)=6$ and $t(W)=1$ by Lemma~\ref{lem:cw_or_acw}, $W$ turns at $0$ as prescribed by $x_0$.)
Note that the definitions of equivalence for pairs $(H,B)$ and triples $(H,\phi,B)$ coincide and, in particular, do not depend on $\phi$.
\QED

\section{From Walks in the Brickwall to Assignment Sets}
\label{sec:assignments}

Throughout Section~\ref{sec:assignments}, $W$ denotes a closed walk in $\BW$ with length $k$.
We first sketch the main idea of this section.
If we consider a solution $(H,\phi,B)$ to $W$, then we mentioned above that this defines which edges of $H$ are horizontal and vertical. Now if we start at a boundary vertex $B_i$ of $H$ that is incident with a horizontal interior edge of $H$, then we can continue following this horizontal path of $H$ until we end in a different boundary vertex $B_j$. We will say that this solution {\em assigns} $i$ to $j$. If we only know all assignments defined by the solution this way, we can reconstruct the unique solution. We will deduce properties of such sets of assignments such that there is a solution if and only if these properties are satisfied. The purpose is to show that we may focus on counting such assignment sets instead of solutions to the walk.

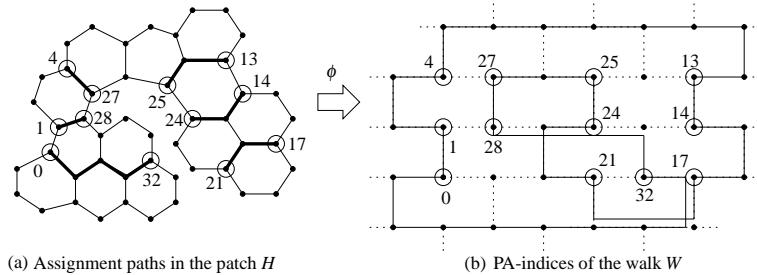
\begin{figure}
\centering
\scalebox{0.7}{$\input{PA-indB.pstex_t}$}
\caption{Assignment paths of a solution and PA-indices of a walk.}
\label{fig:PAindices}
\end{figure}

For all $i,j$ where $W_i$ and $W_j$ lie on the same height, \notat{H_{i,j}} denotes the horizontal walk in $\BW$ from $W_i$ to $W_j$.
Consider an index $i\in \{0,\ldots,k-1\}$ and the vertex $W_i$. Let $u$ be the neighbor of $W_i$ in $\BW$ not equal to $W_{i-1}$ or $W_{i+1}$. 
If $u$ has the same height as $W_i$ and $W$ makes a left turn at $i$, then index $i$ is called a \term{PA-index}. 
In Figure~\ref{fig:PAindices}(b) an example 
is shown, where vertices corresponding to PA-indices are encircled, and their indices are shown.
Note that if $W$ has a 
clockwise
solution $(H,\phi,B)$, then the PA-indices are precisely those indices $i$ such that $B_i$ has degree 3 and the interior edge incident with $B_i$ is horizontal (see Figure~\ref{fig:PAindices}(a)).

A \term{possible assignment (PA)} is a pair $\{i,j\}$ of PA-indices with $W_i\not=W_j$ such that $W_i$ and $W_j$ have the same height and $H_{i,j}$ 
does not contain any of $W_{i-1},W_{i+1},W_{j-1},W_{j+1}$ (note that $H_{i,j}$ has non-zero length). 
For instance, in Figure~\ref{fig:PAindices}(b) some PAs are $\{1,28\}$, $\{1,14\}$ and $\{21,32\}$, but $\{1,24\}$ is not.

Let $(H,\phi,B)$ be a clockwise solution to a closed walk $W$ in $\BW$. An \term{assignment path} $P$ is a horizontal path in $H$ from $B_i$ to $B_j$ where $i\not=j$, and all edges and internal vertices of $P$ are interior edges and vertices of $H$. In Figure~\ref{fig:PAindices}(a) the assignment paths of the given solution are shown in bold.

\begin{propo}[*]
\label{propo:assignments_are_PAs}
If a clockwise solution $(H,\phi,B)$ to $W$ contains an assignment path from $B_i$ to $B_j$, then $\{i,j\}$ is a PA of $W$.
\end{propo}

This motivates the following definition. A clockwise solution $S=(H,\phi,B)$ to a walk $W$ \term{assigns $i$ to $j$} if there is an assignment path from $B_i$ to $B_j$. For each clockwise solution $S$, we define the set
\notat{\AS(S)}$:=\{\{i,j\}:\{i,j\}\mbox{\ is a PA of $W$ and $S$ assigns $i$ to $j$}\}.
$
This is the \term{assignment set} defined by the solution $S$.

\begin{lem}[*]
\label{lem:well-def}
Let $W$ denote a closed walk in $\BW$ and let $S,S'$ be clockwise solutions of $W$. If $S$ and $S'$ are equivalent, then $\AS(S)=\AS(S')$.
\end{lem}

Now we will deduce the properties of a set $\AS(S)$.
Proposition~\ref{propo:apath_partition} shows that assignment paths do not share vertices. Combining this with planarity yields Proposition~\ref{propo:separation}.

\begin{propo}[*]
\label{propo:apath_partition}
Let $(H,\phi,B)$ be a clockwise solution to $W$. Every interior vertex of $H$ and every vertex $B_i$, where $i$ is a PA-index, lies on a unique assignment path. 
\end{propo}

\begin{propo}[*]
\label{propo:separation}
Let $S$ be a solution to a closed walk $W$ that assigns $i$ to $j$. For any $x,y$ with $x<i<y<j$ or $i<x<j<y$, $S$ does not assign $x$ to $y$.
\end{propo}

These two propositions give us properties a set of the form $\AS(S)$ for a clockwise solution $S$ necessarily has to have. Given $W$, a set $A$ of possible assignments of $W$ is a \term{perfect matching} on the set of PA-indices if for every PA-index $i$ of $W$ there is exactly one pair $\{i,j\}\in A$. $A$ is \term{non-crossing} if there do not exist assignments $\{i,j\},\{x,y\}\in A$ such that $i < x < j < y$.
An \term{assignment set} for $W$ is a set of possible assignments of $W$. It is a \term{proper assignment set} if it is a non-crossing, perfect matching on the set of PA-indices of $W$. Combining Proposition~\ref{propo:assignments_are_PAs}, Proposition~\ref{propo:apath_partition} and Proposition~\ref{propo:separation} yields Lemma~\ref{lem:necessary}.
Lemma~\ref{lem:surjective} states more or less the reverse; the long proof appears in Appendix~\ref{sec:surjective}.

\begin{lem}
\label{lem:necessary}
If $S=(H,\phi,B)$ is a clockwise solution of $W$ then $\AS(S)$ is a
proper assignment set for $W$.
\end{lem}

\begin{lem}[*]
\label{lem:surjective}
Let $W$ denote a closed walk in $\BW$ with $t(W)=1$, and let $A$ be a proper assignment set of $W$. Then there exists a clockwise solution $S$ of $W$ with $\AS(S)=A$.
\end{lem}

It remains to establish the converse of Lemma~\ref{lem:well-def}.
Suppose we have two solutions $S=(H,\phi,B)$ and $S'=(H',\phi',B')$ with $\AS(S)=\AS(S')$.
Every vertex of $H$ and $H'$ lies on the boundary or on an assignment path (Proposition~\ref{propo:apath_partition}).
Therefore we can use the boundary and the assignment paths to define a bijection $\psi:V(H)\rightarrow V(H')$. When doing this appropriately, it can be shown that $\psi$ an equivalence. 

\begin{lem}[*]
\label{lem:injective}
Let $W$ be a closed walk in $\BW$, and let $S$ and $S'$ denote clockwise solutions of $W$. If $\AS(S)=\AS(S')$, then $S$ and $S'$ are equivalent.
\end{lem}

\begin{thm}
\label{thm:equivalence}
Let $W$ be a walk in $\BW$ with $t(W)=1$. The number of equivalence classes of solutions to $W$ is the same as the number of different proper assignment sets for $W$.
\end{thm}
\PF
The above lemmas show that $S\mapsto \AS(S)$ gives a bijection from
the set of equivalence classes of clockwise solutions of $W$
to the set of proper assignment sets for $W$, since the following properties are satisfied:
(1) \emph{$\AS$ is well-defined:} Let $S_1$ and $S_2$ denote clockwise solutions of $W$. If $S_1$ and $S_2$ are equivalent, then $\AS(S_1)=\AS(S_2)$ (Lemma~\ref{lem:well-def}).
(2) \emph{The range of $\AS$ is correct:} For any clockwise solution $S$ of $W$ the set $\AS(S)$ is a proper assignment set for $W$ (Lemma~\ref{lem:necessary}).
(3) \emph{$\AS$ is injective:}  Let $S_1$ and $S_2$ denote clockwise solutions of $W$. If $\AS(S_1)=\AS(S_2)$, then $S_1$ and $S_2$ are equivalent (Lemma~\ref{lem:injective}).
(4) \emph{$\AS$ is surjective:} For any proper assignment set $A$ for $W$, there exists a clockwise solution $S$ of $W$ with $\AS(S)=A$ (Lemma~\ref{lem:surjective}).
\QED

It follows that for solving the Hexagonal Patch problem, we may focus on counting proper assignment sets for the walk $W$ (assuming $t(W)=1$).

\section{Counting Maximum Planar Matchings}
\label{sec:circlegraphs}

In this section we will observe that the remaining algorithmic problem is that of counting independent sets in circle graphs, and present a fast algorithm for this problem.
We use the closed walk $W$ in $\BW$ to construct a graph
$\GPA$ with vertex set $V=\{0,\ldots,n-1\}$, where $n$ is the number of PA-indices of $W$. Let $p_0,\ldots,p_{n-1}$ be all PA-indices of $W$, numbered according to their order in $W$. Then the edge set of $\GPA$ will be
$E=\{ij\mid \{p_i,p_j\} \mbox{ is a PA of }W\}$.
The following lemma is now easily observed.

\begin{lem}
\label{lem:equiv_to_MPM}
Let $\GPA$ be the graph as constructed above from the walk $W$.
If $\GPA$ has no perfect planar matching, then $W$ has no proper assignment sets. Otherwise the number of proper assignment sets for $W$ is equal to the number of MPMs in $\GPA$.
\end{lem}

Now we will present an algorithm for counting MPMs of a graph $G$ with $V(G)=\{0,\ldots,n-1\}$. As mentioned in the introduction, this is equivalent to counting maximum independent sets in a circle graph $H$, where $G$ is the chord model graph of $H$.
We will present this algorithm for the general case where $G$ has edge weights: $w_{ij}$ denotes the edge weight of $ij$, and a PM $M$ is {\em maximum} if $\sum_{e\in M} w_e$ is maximum.

For $i,j\in V(G)$ with $i\leq j$, let \notat{G_{i,j}}$=G\left[ \{i,\ldots,j\} \right]$.
If $i>j$, then $G_{i,j}$ is the empty graph.
For $i,j\in V(G)$, let \notat{S_{i,j}} denote the size of a MPM in $G_{i,j}$. In particular, $S_{0,n-1}$ is the size of a MPM in $G$.
We now give a subroutine $S(i,j)$ for calculating $S_{i,j}$, which considers the sizes of various PMs for $G_{i,j}$, and returns the size of the largest PM. 

\begin{tabbing}
\qquad \=\quad \=\quad \= \quad \kill
A subroutine $S(i,j)$ for calculating $S_{i,j}$:\\

(1) \> $m:=0$\\

(2) \> {\bf If} $i<j$ {\bf then} $m:=S_{i+1,j}$\\

(3) \> {\bf For} $v\in N(i)$ with $i+1\leq v\leq j-1$:\\

(4) \> \> $m:=\max\{m,S_{i,v}+S_{v+1,j}\}$\\

(5) \> {\bf If} $j\in N(i)$ and $j>i$ {\bf then}\\

(6) \> \> $m:=\max\{m,w_{ij}+S_{i+1,j-1}\}$\\

(7) \> Return $m$

\end{tabbing}

\begin{lem}[*]
\label{lem:Sroutine}
Let $G$ be a graph with $V(G)=\{0,\ldots,n-1\}$ and $i,j\in V(G)$.
If the values $S_{x,y}$ are known for all $x,y$ with $-1\leq y-x<j-i$, then the subroutine $S(i,j)$ computes $S_{i,j}$ in time $O(d(i))$.
\end{lem}

Let \notat{N_{i,j}} denote the number of MPMs in $G_{i,j}$. 
Below is a similar subroutine $N(i,j)$ for calculating $N_{i,j}$, which considers various PMs for $G_{i,j}$, checks whether they are maximum by comparing the size with $S_{i,j}$, and keeps track of the number of MPMs using the variable $N$.

\begin{tabbing}
\qquad \=\quad \=\quad \= \quad \kill
A subroutine $N(i,j)$ for calculating $N_{i,j}$:\\

(1) \> {\bf If} $j\leq i$ {\bf then} Return 1, {\bf exit}.\\

(2) \> $N:=0$\\

(3) \> {\bf If} $S_{i,j}=S_{i+1,j}$ {\bf then} $N:=N+N_{i+1,j}$\\

(4) \> {\bf For} $v\in N(i)$ with $i+1\leq v\leq j-1$:\\

(5) \> \> {\bf If} $S_{i,j}=S_{i,v}+S_{v+1,j}$ {\bf then}

		$N:=N + N_{i,v}\times N_{v+1,j}$\\

(6) \> {\bf If} $j\in N(i)$ and $S_{i,j}=w_{ij}+S_{i+1,j-1}$ {\bf then}

 $N:=N + N_{i+1,j-1}$\\

(7) \> Return $N$

\end{tabbing}

\begin{lem}[*]
\label{lem:Nroutine}
Let $G$ be a graph with $V(G)=\{0,\ldots,n-1\}$ and $i,j\in V(G)$.
If the values $S_{x,y}$ and $N_{x,y}$ are known for all $x,y$ with $-1\leq y-x< j-i$, and $S_{i,j}$ is known,
then the subroutine $N(i,j)$ computes $N_{i,j}$ in time $O(d(i))$.
\end{lem}

\begin{thm}
\label{thm:MPM}
Let $G$ be a graph with $V(G)=\{0,\ldots,n-1\}$ on $m$ edges.
The size and number of MPMs of $G$ can be computed in time $O(nm)$.
\end{thm}
\PF
For $d=-1$ to $n-1$, we consider all $i,j\in \{0,\ldots,n-1\}$ with $j-i=d$, and calculate $S_{i,j}$ and $N_{i,j}$ using the above subroutines.
This way, for every value of $d$, 
every vertex of $G$ is considered at most once in the role of $i$. For this choice of $i$, calculating $S_{i,j}$ and $N_{i,j}$ takes time $O(d(i))$ (Lemma~\ref{lem:Sroutine}, Lemma~\ref{lem:Nroutine}).
Hence for one value of $d$ this procedure takes time $O(\sum_{i\in V(G)} d(i))=O(m)$.\QED

We remark that Valiente's algorithm~\cite{Val03} for simple (1-regular) chord model graphs can also be extended by using Subroutine $N(i,j)$ to calculate $N_{i,j}$ in constant time, immediately any time after a value $S_{i,j}$ is calculated. This then yields time complexity $O(l)$ and space complexity $O(n)$. In some cases it may be better to transform to a simple chord model graph and use this algorithm.

\section{Summary of the Algorithm}
\label{sec:alg}

We now summarize how counting the number of hexagonal patches that satisfy a given boundary code $X$ of length $k$ can be done in time $O(k^3)$. 
W.l.o.g. $d_2(X)-d_3(X)=6$.
First use $X$ to construct a walk $W$ in $\BW$ of length $k$, as shown in Section~\ref{sec:representations}. Theorem~\ref{thm:equiv_boundary_walk} shows that we may now focus on counting clockwise solutions to $W$. If $W$ is not closed it clearly has no solution.
Since $d_2(X)-d_3(X)=6$ we may now assume $t(W)=1$. Then Theorem~\ref{thm:equivalence} shows we may focus on counting proper assignment sets for $W$. Now construct $\GPA$ as shown in Section~\ref{sec:circlegraphs}. $\GPA$ has $n$ vertices where $n<k$ is the number of PA-indices of $W$ (and $O(n^2)$ edges). By Lemma~\ref{lem:equiv_to_MPM}, the number of proper assignment sets for $W$ is equal to the number of MPMs of $\GPA$, provided that $\GPA$ has a perfect PM. This number and property can be determined in time $O(n^3)\in O(k^3)$ (Theorem~\ref{thm:MPM}).

\section{Discussion}
\label{sec:discussion}

Our first question is whether the complexity of $O(k^3)$ can be improved.
Secondly, considering the motivation from benzenoid hydrocarbons, it is interesting to study whether a patch exists that has a `reasonably flat' embedding in $\mathbb{R}^3$ using {\em regular} hexagons. More precisely, this is the brickwall walk problem, but requires in addition giving a consistent linear order (`depth') for all vertices 
mapped to the same vertex of $\BW$.
It may also be interesting to study generalizations 
such as to surfaces of higher genus. 

After we presented an early version of this work~\cite{BB08}, Jack Graver 
pointed us to a similar well-studied problem in topology. 
Let $S^1$ denote the unit circle and $D^2$ the unit disk in $\mathbb{R}^2$. An {\em immersion} is a continuous function $f:A\rightarrow B$ such that for every $x$ in $A$ there is a neighborhood $N$ of $x$ such that $f|_N$ is a homeomorphism. 
(A {\em curve} when $A=S^1$, $B=\mathbb{R}^2$.)
An immersion $c:S^1\rightarrow \mathbb{R}^2$ of the circle into the plane is {\em normal} if $c$ has only finitely many double-points and $c$ crosses itself at each of these. Two immersions $d$, $d'$ are
{\em equivalent} if there exists a homeomorphism $\phi:\mathbb{R}^2\rightarrow \mathbb{R}^2$ such that $d \circ \phi = d'$.
Now the {\em Immersion Extension} problem is this: given an 
immersion $c:S^1\rightarrow \mathbb{R}^2$, how many immersions $d:D^2\rightarrow \mathbb{R}^2$ exist that extend $c$?  
Note that this problem is not combinatorial, therefore it makes no sense to study its computational complexity. One can turn it into a combinatorial problem by restricting the input to {\em piecewise linear (PL)} curves $c:S^1\rightarrow \mathbb{R}^2$.

When viewing the walk constructed in Section~\ref{sec:representations} as a curve, there are obvious similarities between the Hexagonal Patch problem and the Immersion Extension problem. However, to our knowledge it is an open problem to prove that these problems are in fact equivalent. 
The ideas introduced here may be helpful for giving such a proof.
Establishing this would provide insight to both problems, since
the Immersion Extension problem is well-studied -- at least on normal curves -- see e.g.~\cite{Bla67,EM08,SW92}. 
Interestingly, Blank~\cite{Bla67,Fra70} reduces the Immersion Extension problem problem to a combinatorial problem that is essentially the same as counting MPMs in simple chord model graphs. He does not address the complexity of this problem. Shor and Van Wyk~\cite{SW92} were the first to study the complexity of the combinatorial Immersion Extension problem on normal curves. They give an $O(n^3 \log n)$ algorithm where $n$ is the number of pieces of the PL curve $c$.
Assuming the equivalence of the Immersion Extension problem and the Hexagonal Patch problem, this would give an alternative algorithm for Hexagonal Patch; note that there are methods for transforming general PL curves to equivalent normal PL curves~\cite{Sei98}.
Since our algorithm does not need such a step, it is not only faster but also much easier to implement (see also~\cite{Sei98}).
However, the question of equivalence of these problems is still interesting because many generalizations of the Immersion Extension problem have been studied~\cite{EM08}.
Finally, we believe that in fact our method can be adapted to give a simple and fast 
for the combinatorial Immersion Extension problem that does not require the assumption that the given curve is normal, but that is beyond the scope of this paper.

\noindent
{\bf Acknowledgement} We thank Gunnar Brinkmann for introducing us to this subject and his suggestions, and Hajo Broersma for the discussions on this topic.
%

\input{hexabsIclean.bbl}
\newpage
\appendix

\section{Proofs of Section~\ref{sec:representations} and Section~\ref{sec:assignments}}

{\bf Proof} of Lemma~\ref{lem:cw_or_acw}:
Let $\phi$ be a LIH from a hexagonal patch $H$ with boundary cycle $B$ of length $k$ to $\BW$, and let $W=\phi(B_0),\phi(B_1),\ldots,\phi(B_{k-1}),\phi(B_0)$.
We first show that $(H,\phi,B)$ is either a clockwise solution to the walk $W$ or an anticlockwise solution to $W$.

An index $i$ is called {\em locally clockwise} if either $d(B_i)=2$ and $W$ makes a right turn at $i$, or $d(B_i)=3$ and $W$ makes a left turn at $i$.

We show that if some $i$ is locally clockwise, then every index is locally clockwise. Suppose this is not true, so then there is an $i$ that is locally clockwise such that $i+1$ is not.
Assume first that $d(B_i)=2$ and $d(B_{i+1})=2$. Then $W$ makes a right turn at $i$, but a left turn at $i+1$. Therefore $W_{i-1}$ and $W_{i+2}$ do not lie at a common facial cycle of $\BW$. Since $B_i$ and $B_{i+1}$ both have degree 2 in $H$, all of the vertices $B_{i-1},\ldots,B_{i+2}$ lie at a common inner face of $H$. This is a contradiction since $\phi$ maps faces of $H$ to faces of $\BW$.

In the case where $d(B_i)=d(B_{i+1})=3$, we consider the neighbor $v_i$ of $B_i$ that is not equal to $B_{i-1}$ or $B_{i+1}$, and the neighbor $v_{i+1}$ of $B_{i+1}$ that is not equal to $B_i$ or $B_{i+2}$. These again lie at a common face of $H$, but if $W$ makes a left turn at $i$ and a right turn at $i+1$, are mapped to two vertices that do not lie at a common face, which again yields a contradiction. The two other cases are analogous.
We conclude that if a solution contains a locally clockwise vertex, it is clockwise.

Now we relate this to the turning number.
Let $d_i$ denote the number of vertices of degree $i$ on the boundary of $H$. If $(H,\phi,B)$ is a clockwise solution then $d_2=\convex(W)$ and $d_3=\concave(W)$. 
We know that $d_2-d_3=6$ since $H$ contains no 5-faces. Hence $t(W)=1$. Similarly, if an anticlockwise solution exists then $t(W)=-1$ follows, which proves the statement.\QED

{\bf Proof} of Proposition~\ref{propo:assignments_are_PAs}:
We show that if a clockwise solution $(H,\phi,B)$ to $W$ assigns $B_i$ to $B_j$, then $\{i,j\}$ is a PA of $W$.
Let $P$ be an assignment path from $B_i$ to $B_j$. We have that $B_i$ and $B_j$ have the same height, since $P$ is horizontal. 
$B_i\not=B_j$ holds since $H$ is 2-connected. $P$ is then mapped to a non-zero length path in $\BW$ (it does not turn back, since $\phi$ is a LIH), so $W_i\not=W_j$ follows.
All edges of $P$ are interior edges of $H$, so $B_i$ and $B_j$ have degree 3, and therefore $W$ makes a left turn at $i$ and $j$. It follows that
$i$ and $j$ are PA-indices.
Since $\phi$ is a LIH and $P$ contains no boundary edges, $H_{i,j}$ does not contain any of $W_{i-1},W_{i+1},W_{j-1},W_{j+1}$.
\QED

{\bf Proof} of Lemma~\ref{lem:well-def}:
We show that if two clockwise solutions $S$ and $S'$ of $W$ are equivalent, then $\AS(S)=\AS(S')$.
Let $S=(H,\phi,B)$ and $S'=(H',\phi',B')$, and let $\psi:V(H)\rightarrow V(H')$ demonstrate their equivalence. 
Note that $\phi'\circ\psi$ and $\phi$ are both LIHs from $H$ into $\BW$ that map $B_i$ to $W_i$. As the LIH with this property is uniquely determined by Lemma~\ref{lem:unique_phi}, we conclude that $\phi'\circ\psi=\phi$.
In particular, any edge $uv$ is horizontal in $H$ if and only if $\psi(u)\psi(v)$ is horizontal $H'$. Clearly an analogous statement holds for vertices being interior. Therefore $\psi$ maps assignment paths to assignment paths. Since $\psi(B_i)=B'_i$ for all $i$ it follows that $\AS(S)=\AS(S')$.
\QED

{\bf Proof} of Proposition~\ref{propo:apath_partition}:
Let $(H,\phi,B)$ be a clockwise solution to $W$. We show that every interior vertex of $H$ and every vertex $B_i$, where $i$ is a PA-index, lies on a unique assignment path.
Let $M$ be the set of horizontal non-boundary edges of $H$, 
and let $H'=(V(H),M)$. 
Since $\phi$ is a LIH, $H'$ has maximum degree at most 2. $H'$ contains no cycles, because these would have to be mapped to cycles of $\BW$ but $\BW$ contains no cycles with only horizontal edges. Hence $H'$ is a set of paths and isolated vertices.

It can be seen that vertices with degree 2 in $H'$ are interior vertices of $H$, and that vertices with degree 1 in $H'$ are equal to $B_i$ for some PA-index $i$.
Hence the path components of $H'$ (paths of non-zero length)
are all assignment paths. 
Since all assignment paths in $H$ are also part of $H'$, we see that there is a one-to-one correspondence between assignment paths in $H$ and non-trivial components of $H'$. 
We also see that every interior vertex of $H$ and every vertex $B_i$ where $i$ is a PA-index lies on one such path. The statement follows.\QED

{\bf Proof} of Proposition~\ref{propo:separation}:
We show that if a solution $(H,\phi,B)$ assigns $i$ to $j$ and $i<x<j<y$, then it does not assign $x$ to $y$.
Suppose $H$ contains an assignment path $P$ from $B_i$ to $B_j$, and an assignment path $Q$ from $B_x$ to $B_y$. 
By Proposition~\ref{propo:apath_partition}, $P$ and $Q$ have no vertices in common.
But since the (distinct) end vertices of the paths appear in the order $B_i$, $B_x$, $B_j$ and $B_y$ along a boundary cycle of the plane graph $H$, this is impossible.
(Formally, to obtain a contradiction, we may use $P$, $Q$ and the boundary cycle of $H$ to exhibit a subdivision of $K_4$ that is embedded with all vertices on the boundary, which then would yield a planar embedding of $K_5$.)
\QED

For the proof below and later proofs in the appendix, it is important to distinguish between two different kinds of horizontal edges of $\BW$: \term{horizontal left} (\term{horizontal right}) edges are edges that follow a vertical edge after turning left (right). Note that this partitions the edges of $\BW$ into vertical edges, horizontal left edges, and horizontal right edges, and that every face contains two of each. The same holds for faces in a solution $(H,\phi,B)$;
Recall that if patch $H$ is mapped by a LIH $\phi$ to $\BW$, this allows us to define vertical and horizontal (left / right) edges in $H$. Similarly, we will talk about vertices of $H$ that lie to the left / below etc.\ other vertices. This is also defined by $\phi$ and the chosen drawing of $\BW$.

\medskip

{\bf Proof} of Lemma~\ref{lem:injective}:
We show that if $\AS(S)=\AS(S')$ for two solutions $S$ and $S'$, then these solutions are equivalent.
Let $S=(H,\phi,B)$ and $S'=(H',\phi',B')$.
We construct the isomophism $\psi$ from $H$ to $H'$ that will demonstrate the equivalence as follows. For all $i$, $\psi(B_i)=B'_i$. This defines $\psi$ for boundary vertices. Every non-boundary vertex lies on a unique assignment path (Proposition~\ref{propo:apath_partition}). Suppose such a vertex $v$ lies on an assignment path $P$ from $B_i$ to $B_j$. Then $\{i,j\}\in \AS(S)$ and thus $\{i,j\}\in \AS(S')$. The assignment path $P'$ from $B'_i$ to $B'_j$ in $H'$ is also mapped by $\phi'$ to $H_{i,j}$ and therefore has the same length as $P$ (since $\phi$ and $\phi'$ are LIHs). Now if $v$ is the $x$-th vertex on $P$, $\psi$ will map $v$ to the $x$-th vertex of $P'$. This defines $\psi$. Since every vertex of $H$ lies on the boundary or on an assignment path, the function $\psi$ is defined for every vertex of $H$, and since the same holds for $H'$, $\psi$ is a bijection. By definition $\psi$ maps boundary vertices to the correct boundary vertices, so to demonstrate that $\psi$ is an equivalence between $S$ and $S'$, it only remains to show that it is an isomorphism.

We only show that edges of $H$ are mapped to edges of $H'$ by $\psi$. By symmetry a similar statement then follows for $\psi^{-1}$, which proves that $\psi$ is an isomorphism. Clearly $\psi$ maps boundary edges of $H$ to boundary edges of $H'$. Observe that every horizontal non-boundary edge of $H$ 
lies on an assignment path. Therefore $\psi$ also maps horizontal edges of $H$ to edges of $H'$. What remains are vertical edges of $H$ that do not lie on the boundary. Note that $\psi$ maps horizontal left (right) edges of $H$ to edges of $H'$ of the same type, and by observing the same for $\psi^{-1}$, it also follows that if a vertical edge is mapped to an edge, it is mapped to a vertical edge again.

Suppose there exists a (vertical, interior) edge of $H$ that is not mapped to an edge of $H'$ by $\psi$. Let $e=u_1v_1\in E(H)$ be such an edge such that all edges that lie to the left of it
are mapped to edges of $H'$. Suppose $u_1$ lies below $v_1$.
$e$ is incident with two inner faces of $H$, so we may choose $F=u_1,u_2,u_3,v_3,v_2,v_1,u_1$ to be the inner face of $H$ on the left side of $e$. Note that all other edges of $F$ are mapped to edges of $H'$; four edges are horizontal, and the other vertical edge is mapped by our choice of $e$.

Let $u'_1$, $u'_2$, $u'_3$, $v'_3$, $v'_2$ and $v'_1$ respectively be the images under $\psi$ of the vertices of $F$. 
The edges $v'_1v'_2$ and $v'_2v'_3$ lie on a common face of $H'$ (since $v'_2$ has degree at most 3). We show that they lie on a common {\em inner} face of $H'$. If not, then both $v'_1v'_2$ and $v'_2v'_3$ are boundary edges. Then the corresponding edges $v_1v_2$ and $v_2v_3$ of $H$ are boundary edges too, and thus these two edges share both an inner face and the outer face. Since $H$ is 2-connected, it follows that $d(v_2)=2$. Because both $H$ and $H'$ are clockwise solutions, $d(v'_2)=2$. Hence the two edges in $H'$ also share two faces, and thus one inner face.

Let $F'=v'_1,v'_2,v'_3,x,y,z,v'_1$ be the inner face of $H'$ on which these two edges lie. Since $\psi$ maps to edges of the same type, $v'_1v'_2$ and $v'_2v'_3$ are horizontal left and horizontal right edges respectively.
Since $\phi'$ maps inner faces of $H'$ to inner faces of $\BW$, $v'_3x$ is a vertical edge. $v'_3$ is incident with at most one vertical edge ($\phi'$ is a LIH), so we may conclude that $x=u'_3$ (here we use the fact that vertical edges are not mapped to horizontal edges, so $v'_3u'_3$ is vertical). Continuing this reasoning shows that $F'=v'_1,v'_2,v'_3,u'_3,u'_2,u'_1,v'_1$. Hence $u'_1v'_1\in E(H')$, a contradiction with the choice of $u_1v_1$. We conclude that $\psi$ is an isomorphism, which concludes the proof.\QED

\section{The proof of Lemma~\ref{lem:surjective}}
\label{sec:surjective}

Before we can prove Lemma~\ref{lem:surjective} we need to introduce some new terminology and lemmas.
For a closed walk $W=v_0,\ldots,v_k$, the {\em subwalk of $W$ from $i$ to $j$} is the walk $v_i,v_{i+1},\ldots,v_j$ of length $j-i \bmod k$. 
If $j<i$ then, more precisely, this is the walk $v_i,v_{i+1},\ldots,v_{k-1},v_0,\ldots,v_j$.
The subwalk of $W$ from $i$ to $j$ will be denoted by \notat{W_{i,j}}.
If $W=v_0,\ldots,v_k$ is a closed walk, then for any $i\in \{0,\ldots,k-1\}$, the walk $W'=v_i,v_{i+1},\ldots,v_{k-1},v_0,v_1,\ldots,v_{i-1},v_i$ is called a \term{rotation} of $W$. We will write \notat{W'\approx W} to express that $W'$ is a rotation of $W$.
For a pair of walks $W=v_0,\ldots,v_k$ and $W'=u_0,\ldots,u_l$ with $v_k=u_0$ and $v_{k-1}\not=u_1$, \notat{W\circ W'} denotes the \term{concatenation} of $W$ and $W'$, which is $v_0,\ldots,v_k,u_1,\ldots,u_l$.
In sequences, the notation \notat{(a)^b} means that $b$ copies of $a$ are inserted in the sequence at this point. For instance, $1,2,(3)^3,4$ denotes the sequence $1,2,3,3,3,4$.

Let $W$ be a closed walk in $\BW$.
For indices $i$, let \notat{n(i)} be the first PA-index after $i$ (not equal to $i$). So with respect to the walk $W$ shown in Figure~\ref{fig:PAindices}(b), $n(1)=4$, $n(32)=0$, etc.
The subwalks of the form $W_{i,n(i)}$ for any PA-index $i$ are called the \term{pieces} of $W$.

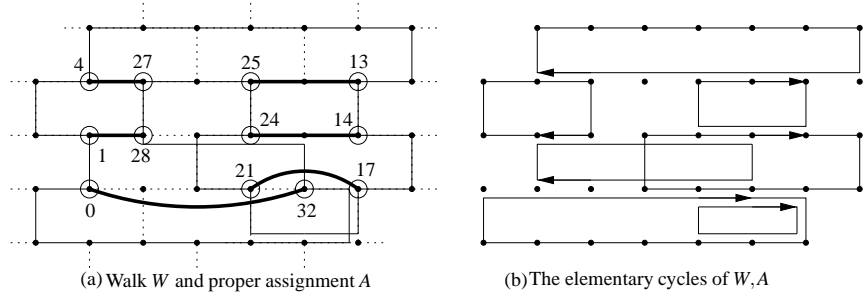
\begin{figure}
\centering
\scalebox{0.75}{$\input{elcycsB.pstex_t}$}
\caption{Elementary cycles in $\BW$ given by a proper assignment $A$ of $W$.}
\label{fig:elcycs}
\end{figure}

\paragraph{Elementary Cycles}
Let $A$ be a proper assignment set of a closed walk $W$ in $\BW$. 
We will now define how such a tuple $W,A$ gives {\em elementary cycles}, which correspond to closed walks in $\BW$. Informally, for any index $i$ the unique elementary cycle that contains $W_iW_{i+1}$ can be found as follows. This is illustrated in Figure~\ref{fig:elcycs}, where the proper assignment given by the solution shown in Figure~\ref{fig:PAindices}(a) is used. Arcs are shown to indicate the direction of the elementary cycles, and the head of the arc indicates the first vertex.
 
Start at $W_i$, and continue along $W$ (in the direction of increasing indices) until the first PA-index $n(i)$ is found. Let $\{n(i),i_1\}\in A$ be the (unique) assignment in $A$ that contains $n(i)$. Then continue along the horizontal path $H_{n(i),i_1}$ in $\BW$. At $W_{i_1}$, follow $W$ again until the next PA-index $n(i_1)$, and then follow $H_{n(i_1),i_2}$ where $\{n(i_1),i_2\}\in A$, etc. Continue with this procedure, taking alternatingly pieces of $W$ and horizontal paths that correspond to assignments in $A$ until we arrive again at $W_i$. 
Observe that this procedure ends and that we do actually arrive at $W_i$ again this way. We consider the choice of starting vertex to be irrelevant, hence in the following formal definition we fix a canonical rotation.
An \term{elementary cycle} of $W,A$ is a walk in $\BW$ of the form
\[
C=W_{i_0,n(i_0)} \circ H_{n(i_0),i_1} \circ W_{i_1,n(i_1)} \circ \ldots \circ H_{n(i_l),i_0},
\]
where $\{n(i_j),i_{j+1}\}\in A$ for all $j$, and $\{n(i_l),i_0\}\in A$. In addition, we require that $i_0<i_j$ for all $j\in \{1,\ldots,l\}$. This last condition fixes the canonical rotation. 
Note that elementary cycles that start at different indices of $W$ may still yield the same walk in $\BW$.
Therefore we also consider a sequence $I$ of numbers that give the corresponding indices of $W$. We insert $-1$ in this sequence for the vertices that correspond to vertices of horizontal paths instead of parts of $W$. For example, for the above choice of $C$,
\[
I\ =\ i_0,\ldots,n(i_0)\ ,\ (-1)^{x-1}\ ,\ i_1,\ldots,n(i_1)\ ,\ \ldots\ldots\ ,\ (-1)^{y-1}\ ,\ i_0,
\]
where $x$ and $y$ are the length of $H_{n(i_0),i_1}$ and $H_{n(i_l),i_0}$ respectively. 
 (Obviously, if $i_0>n(i_0)$, then $i_0,\ldots,n(i_0)$ should be read as $i_0,i_0+1,\ldots,k-1,0,1,\ldots,n(i_0)$, where $k$ is the length of $W$, etc.)

Formally, an elementary cycle is now a pair $C,I$ of a closed walk $C$ in $\BW$ and sequence of numbers $I$ that are of the form explained above. 
This formal definition is needed to clearly define what the {\em number} of elementary cycles of $W,A$ is: elementary cycles are still considered different even if they yield the same walk, but admitting different rotations is irrelevant.
However, below we will often informally denote elementary cycles just by $C$; the index sequence $I$ is clear from how we denote $C$. 
Note that for every $i$ there is a unique elementary cycle $C,I$ such that $I$ contains $i$ and $i+1$ consecutively. In a slight abuse of notation, from now on we will often simply express this statement as follows: there is a unique \elcyc\ $C$ that contains the walk edge $W_iW_{i+1}$.
In the special case where $A=\emptyset$, there is only one elementary cycle $C,I$, which has $C=W$.

\medskip

A walk in $\BW$ in which every vertical edge is followed by a horizontal right edge and preceded by a horizontal left edge is called a \term{right-turn walk}.

\begin{propo}
\label{propo:elcyc_rightturn}
Let $A$ be a proper assignment set for a walk $W$ in $\BW$. Then every elementary cycle $C$ of $W,A$ is a right-turn walk.
\end{propo}
\PF
Any vertical edge of $C$ must come from a piece of $W$. So let $W_{i-1}W_{i}$ be this vertical edge. If $W$ turns left at $i$, then $i$ is a PA-index and $C$ turns right. If $W$ turns right at $i$, then $i$ is not a PA-index and $C$ turns right as well. Hence in both cases, a vertical edge in $C$ is followed by a horizontal right edge.
Similarly, $C$ turns right at $i-1$ in both the case that it is a PA-index and the case that it is not, hence vertical edges in $C$ are preceded by horizontal left edges.
\QED

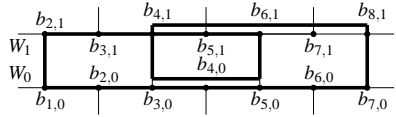
\begin{figure}
\centering
\scalebox{0.75}{$\input{rightturn.pstex_t}$}
\caption{A right-turn walk $W$ in $\BW$ with $t(W)=2$.}
\label{fig:rightturn}
\end{figure}

For the following proof, 
we use the following vertex labelling for $\BW$. See also Figure~\ref{fig:rightturn}, which illustrates the next Proposition.
\[
V(\BW)=\{b_{i,j}:i,j\in \mathbb{Z}\}
\]
\[
E(\BW)=	\{b_{i,j}b_{i+1,j}:i,j\in \mathbb{Z}\} \cup 
	\{b_{i,j}b_{i+1,j+1}:i,j\in \mathbb{Z}, i\mbox{ odd}\}
\]

\begin{propo}
\label{propo:turnnum_closedrightturnwalk}
Let $W$ be a closed right-turn walk in $\BW$. Then $t(W)\geq 1$. If $t(W)=1$, then $W$ has a clockwise solution.
\end{propo}

\PF
For a closed walk $W$ in $\BW$ of length $k$ and any $l\in \{0,\ldots,k\}$ we define $t(W,l)$ to be the number of indices $i$ with $1\leq i\leq l$ such that $W$ makes a right turn at $i$ minus the number of those indices where $W$ makes a left turn. So $t(W,0)=0$ and $t(W,k)=6t(W)$, and for every $l$, $t(W,l+1)=t(W,l)\pm 1$.

Every closed walk in $\BW$ contains a vertical edge (since walks do not turn back), so w.l.o.g. assume $W_0W_1=b_{1,0}b_{2,1}$. Since $W$ is a right-turn walk, after this a horizontal right edge follows, which is part of an alternating sequence of horizontal right and left edges. This sequence continues until a horizontal left edge is followed by a vertical edge, and after that an alternating sequence of horizontal right and left edges again follows (note that the walk cannot close before this point).
So for some $i\geq 1$ and $j\geq 1$ we have 
\[
W_{0,2i+2j+2}=b_{1,0},b_{2,1},b_{3,1},\ldots,b_{2i+2,1},b_{2i+1,0},b_{2i,0},\ldots,b_{2i-2j+1,0}.
\]
(In Figure~\ref{fig:rightturn}, $i=2$ and $j=1$. Note that in general $j>i$ is also possible.)
Choose $j$ maximum, so either the walk $W$ closes at this point ($k=2i+2j+2$), or $W$ continues with another vertical edge.
In either case, the sequence $t(W,0),t(W,1),\ldots,t(W,2i+2j+2)$ is then $0,(1,2)^i,3,(4,5)^j,6$.
Continuing this reasoning shows that $t(W,l)$ can never decrease below 6 when $l\geq i+j+2$, so we conclude $t(W)=t(W,k)/6\geq 1$. In addition, if $t(W)=1$, then the walk cannot contain another vertical edge, so
\[ W=b_{1,0},b_{2,1},b_{3,1},\ldots,b_{2i+2,1},b_{2i+1,0},b_{2i,0},\ldots,b_{1,0}.
\]
It is easily seen that in this case the subgraph of $\BW$ induced by the vertices of $W$ is a clockwise solution to $W$.
\QED

\paragraph{Splitting a walk}
\begin{figure}
\centering
\scalebox{0.75}{$\input{splitB.pstex_t}$}
\caption{Splitting the walk $W$ and assignment set $A$ into two.}
\label{fig:split}
\end{figure}
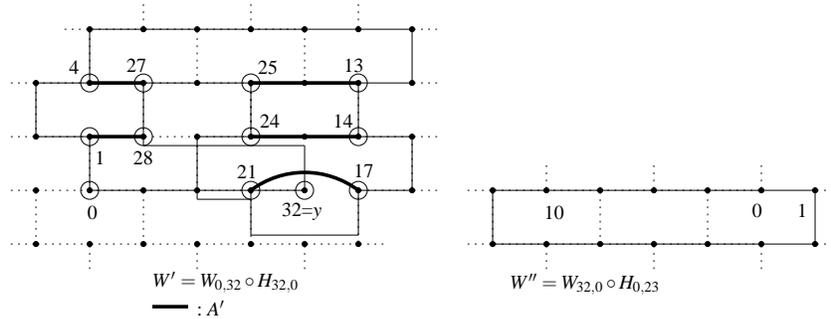
In the following lemmas we use the following notations. Let $A$ again be a proper assignment set for the closed walk $W$ in $\BW$. 
We assume $A$ contains a PA $\{0,y\}$ such that $n(y)=0$ (i.e. $W_{y,0}$ is a piece of $W$).
We now consider the two new closed walks \notat{W'}$=W_{0,y}\circ H_{y,0}$ and \notat{W''}$=W_{y,0}\circ H_{0,y}$ in $\BW$. Note that $W''$ is in fact an \elcyc\ of $W,A$, and that for all $i\in \{0,\ldots,y\}$, $W'_i=W_i$. Figure~\ref{fig:split} illustrates this for the walk and assignment from Figure~\ref{fig:elcycs} (note that $y=32$).

We remark that although we consider various walks in $\BW$ in this section, the notations $H_{i,j}$ and $n(i)$ are always defined with respect to $W$.
Let
\notat{A'}$= A-\{0,y\}$. 
(We denote $A\bs\{x\}$ and $A\cup \{x\}$ as $A-x$ and $A+x$ respectively.)
Observe that $i$ is a PA-index of $W'$ if and only if it is a PA-index of $W$ and $i\not=0,y$. Thus it is easily seen that:
\begin{propo}
\label{propo:newASisproper}
$A'$ is a proper assignment set for $W'$.
\end{propo}

The above construction of smaller walks and corresponding assignment sets from a given pair $W,A$ allows various induction proofs, justified by the next lemma.

\begin{lem}
\label{lem:inductionable}
Let $W$, $A$, $W'$ and $A'$ be as defined above. 
Then
\begin{enumerate}
\item
For every \elcyc\ $C',I'$ of $W',A'$, there is an \elcyc\ $C,I$ of $W,A$ with $C\approx C'$.
\item
$W,A$ has one more \elcyc\ than $W',A'$.
\end{enumerate}
\end{lem} 
\PF
We use the notation $n'$ and $H'_{i,j}$ for the walk $W'$, which are similar to the notations $n$ and $H_{i,j}$ for $W$, so $n'(i)$ denotes the next PA-index of $W'$ after $i$, and $H'_{i,j}$ denotes the horizontal path between $W'_i$ and $W'_j$.
Let $C',I'$ be an \elcyc\ of $W',A'$ with
\[
C'=W'_{i_0,n'(i_0)} \circ H'_{n'(i_0),i_1} \circ W'_{i_1,n'(i_1)} \circ \ldots \circ H'_{n'(i_l),i_0}.
\]
All PAs in $A'$ also appear in $A$, and all PA-indices of $W'$ are PA-indices of $W$. It follows that $C',I'$ is also an \elcyc\ of $W,A$, unless one of the pieces, say $W'_{i_l,n'(i_l)}$ contains the part of $W'$ corresponding to $H_{y,0}$. More precisely, this happens when $n(i_l)=y$ and $n'(i_l)=n(0)$. 
(In Figure~\ref{fig:split}, $n'(28)=1=n(0)$ but $n(28)=32=y$.)
But in that case, we may replace $W'_{i_l,n'(i_l)}$ with $W_{i_l,y} \circ H_{y,0} \circ W_{0,n(0)}$, and choose the appropriate rotation (starting with $0$), which yields an elementary cycle $C,I$ of $W,A$ with $C\approx C'$. This proves the first statement.

The above construction maps \elcyc s of $W',A'$ to \elcyc s of $W,A$. It is easy to see that they are all mapped to different \elcyc s, which only contain pieces that are subwalks of $W_{0,y}$, and that every \elcyc\ of $W,A$ that contains a piece of $W_{0,y}$ is covered this way. It remains to consider \elcyc s of $W,A$ that contain a piece of $W_{y,0}$. Since $\{0,y\}\in A$ and $n(y)=0$, there is only one such \elcyc\ $C,I$ (with $C=W''$, as defined above). This proves the second statement.\QED

As a first application of Lemma~\ref{lem:inductionable}, we can determine the number of \elcyc s.

\begin{corol}
\label{cor:num_elcycs}
Let $A$ be a proper assignment set for closed walk $W$ in $\BW$, and let $\npa$ be the number of PA-indices of $W$. Then the number of elementary cycles of $W,A$ is $\npa/2+1$.
\end{corol}
\PF
If $A=\emptyset$ then there are no PA-indices, and $W$ itself is the only elementary cycle, which proves the statement. 

Otherwise we can use induction: choose a PA $\{x,y\}\in A$ with $x=n(y)$. Such a PA exists since $A$ is a non-crossing perfect matching on the PA-indices. W.l.o.g we may assume that $x=0$, since considering a different rotation of $W$ and changing $A$ accordingly does not change $\npa$ or the number of \elcyc s.
 
Now consider $W'$, $A'$ and $W''$ as defined above using $\{0,y\}$.
Let $\npa'$ denote the number of PA-indices of $W'$.
$W''$ is an \elcyc\ of $W$, hence a right-turn walk, which has no PA-indices.
Therefore every PA-index of $W$ other than $0$ or $y$ corresponds to a PA-index in $W'$, so $\npa'+2=\npa$.
By Lemma~\ref{lem:inductionable} the number $n$ of \elcyc s in $W$ is equal to $n'+1$, where $n'$ is the numbers of \elcyc s of $W',A'$. 
Since $A'$ is a proper assignment set for $W'$ (Proposition~\ref{propo:newASisproper}) we may use induction, so $n=n'+1=\npa'/2+2=\npa/2+1$.
\QED

The above corollary will now be used to deduce that all elementary cycles defined by a {\em proper} assignment set have turning number 1, and hence admit a clockwise solution by Proposition~\ref{propo:turnnum_closedrightturnwalk}.

\begin{lem}
\label{lem:key_counting}
Let $A$ be a proper assignment set for a closed walk $W$ in $\BW$. Then $t(W)=1$ if and only if every elementary cycle $C$ of $A,W$ has $t(C)=1$
\end{lem}

\PF
The number of PA-indices of $W$ is denoted by $\npa$.
The number of left turns that $W$ makes at non-PA-indices is denoted by $\concave^*(W)$. Since $W$ makes a left turn at every PA-index, we have \[
6t(W)=\convex(W)-\concave(W)=\convex(W)-\npa-\concave^*(W).
\]
Let $\EC$ denote the set of all elementary cycles of $A,W$.
When summing the difference between right and left turns over all elementary cycles we obtain
\[
\sum_{C\in\EC} (\convex(C)-\concave(C))=\convex(W)-\concave^*(W)+2\npa.
\]
Here we used the following observations. (i) Elementary cycles make right turns at PA-indices of $W$, and every PA-index contributes a right turn to two elementary cycles. (ii) Non-PA-indices of $W$ contribute the same type of turn to one elementary cycle. (iii) All indices of elementary walks that do not correspond to pieces of $W$ correspond to internal vertices of assignment paths; these vertices of assignment paths contribute a left turn to one elementary cycle and a right turn to another, hence these terms cancel.
Combining this with $|\EC|=\npa/2+1$ (Corollary~\ref{cor:num_elcycs}) yields
\[
6t(W)+3\npa=\convex(W)-\concave^*(W)+2\npa=
\]
\[
\sum_{C\in\EC} (\convex(C)-\concave(C))=\sum_{C\in\EC} (\convex(C)-\concave(C)-6)+3\npa+6 \Longleftrightarrow
\]
\[
t(W)=\sum_{C\in\EC} (t(C)-1)+1.
\]
Because $t(C)-1\geq 0$ for all $C\in \EC$ (Proposition~\ref{propo:elcyc_rightturn}, Proposition~\ref{propo:turnnum_closedrightturnwalk}), this proves the statement.
\QED

\medskip

The proof of the next lemma is illustrated in Figure~\ref{fig:SGL} (using the same example as before).
\begin{figure}
\centering
\scalebox{0.9}{$\input{glueB.pstex_t}$}
\caption{Combining two partial solutions.}
\label{fig:SGL}
\end{figure}
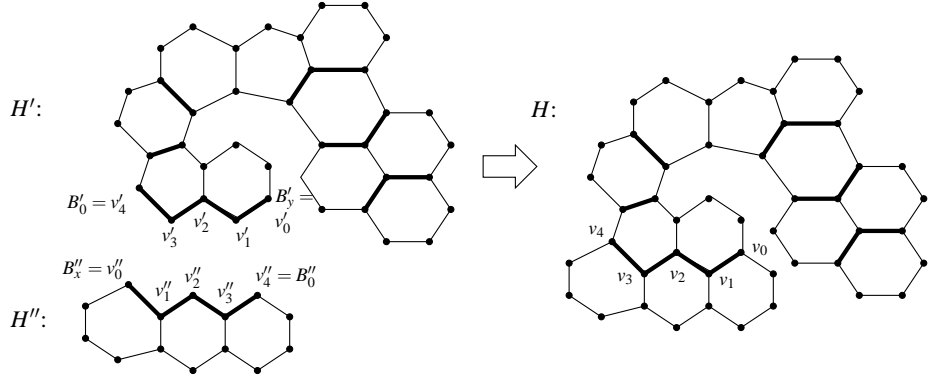
\begin{lem}
\label{lem:gluing-lemma}
Let $W$, $y$, $W'$ and $W''$ be as defined above.
If $S'$ and $S''$ are clockwise solutions of $W'$ and $W''$ respectively, then a clockwise solution $S$ of $W$ exists with 
$\AS(S)=\AS(S')+\{0,y\}$
\end{lem}
\PF
Let $S'=(H',\phi',\Bo)$ be a clockwise solution to $W'=W_{0,y}\circ H_{y,0}$, and $S''=(H'',\phi'',\Bt)$ be a clockwise solution to $W''=W_{y,0}\circ H_{0,y}$. 
Let $x=k-y$ (recall that $k$ is the length of $W$), so $W''_{0,x}=W_{y,0}$.

Let $\Bo_{y,0}=v'_0,v'_1,\ldots,v'_l$, and let $\Bt_{x,0}=v''_0,v''_1,\ldots,v''_l$.
Note that these paths indeed have the same length $l$ and
\begin{equation}
\label{eq:phi}
\phi'(v'_i)=\phi''(v''_{l-i}) \mbox{\ for\ } i=0,\ldots,l.
\end{equation}
Since $\{0,y\}$ is a PA of $W$, we have $l\geq 1$.
$0$ and $y$ are PA-indices, so $W$ makes left turns at $0$ and $y$, and therefore $W'$ makes right turns at $0$ and $y$, and $W''$ makes right turns at $0$ and $x$. Because both $S'$ and $S''$ are clockwise solutions, $d(v'_0)=d(v'_l)=2$ and $d(v''_0)=d(v''_l)=2$ follows.
Since these two paths are mapped to the same paths in $\BW$ but in reverse direction, and both $S'$ and $S''$ are clockwise solutions, we see that $v'_i$ has degree 2 when $v''_{l-i}$ has degree 3 and vice versa, for $i=1,\ldots,l-1$.

Construct $H$ by starting with a copy of $H'$ and a copy of $H''$, and for all $i\in \{0,\ldots,l\}$, identify the vertex $v'_i$ of $H'$ with the vertex $v''_{l-i}$ of $H''$. Call the resulting vertex $v_i$.
For $i\in \{0,\ldots,l-1\}$, this results in two parallel edges between $v_i$ and $v_{i+1}$. Delete one edge of every such parallel pair. 
Because $l\geq 1$, this gives again a 2-connected graph.
Using the above observations on the degrees of $v'_i$ and $v''_i$, we see that $d(v_i)=3$ for all $i$: either a vertex of degree 2 is identified with a vertex of degree 3 and two incident edges are removed (when $i=1,\ldots,l-1$), or two vertices of degree 2 are identified and one incident edge is removed (when $i=0,l$).

Choose an embedding of $H$ in which every inner face of $H'$ or $H''$ is again an inner face of $H$, and which has boundary cycle $v_l,\Bo_{1},\ldots,\Bo_{y-1},v_0,\Bt_{1},\ldots,\Bt_{x-1},v_l$.
A LIH $\phi$ from $H$ is constructed by setting $\phi(u)=\phi'(u)$ for all $u\in V(H)\cap V(H')$ and $\phi(u)=\phi''(u)$ for all $u\in V(H)\cap V(H'')$, and $\phi(v_i)=\phi(v'_i)=\phi(v''_{l-1})$ for the new vertices (see Equation~(\ref{eq:phi})).

We observe that $H$ is a hexagonal patch: 
all inner faces of $H$ correspond to inner faces of $H'$ or $H''$, and thus all have length 6. The degree constraints hold since the new vertices $v_i$ have $d(v_i)=3$ and boundary vertices of $H'$ and $H''$ that were not identified remain boundary vertices.

Next we show that $\phi$ is a LIH. For vertices of $H$ that have no neighbor in $V(H')\bs\{v'_0,\ldots,v'_l\}$ or no neighbor in $V(H'')\bs \{v''_0,\ldots,v''_l\}$, the local injectivity follows from the local injectivity of $\phi'$ and $\phi''$. 
The only two vertices of $H$ for which this does not hold are $v_0$ and $v_l$. But the three neighbors of $v_l$ in $H$ are mapped to $W_1$, $W_{k-1}$ and a vertex on $H_{0,y}$, which are all different. A similar statement holds for $v_0$. This proves that $\phi$ is again a LIH.

We observed above that the boundary cycle of $H$ is 
\[
B=v_l,\Bo_{1},\ldots,\Bo_{y-1},v_0,\Bt_{1},\ldots,\Bt_{x-1},v_l,
\] 
which is mapped by $\phi$ exactly to $W_0,W_{1},\ldots,W_{y-1},W_y,W_{y+1},\ldots,W_{k-1},W_0=W$. 
So $(H,\phi,B)$ is a solution to $W$. Since we started with clockwise solutions, it is also obviously a clockwise solution.
(By Lemma~\ref{lem:cw_or_acw} we do not have to check this for all boundary vertices.)

An assignment path from $\Bo_i$ to $\Bo_j$ in $H'$ yields an assignment path from $B_{i}$ to $B_{j}$ in $H$.
In addition, the path $v_0,\ldots,v_l$ in $H$ is mapped by $\phi$ to $H_{y,0}$, and no internal vertices of this path lie on the boundary of $H$. Hence this solution assigns $0$ to $y$. Now all PA-indices of $W$ are accounted for ($W''$ contains no PA-indices), which shows that 
$\AS(S)=\AS(S')+\{0,y\}$.
\QED

Finally, we have collected all the ingredients that are necessary to prove the remaining lemma.

\medskip

{\bf Proof} of Lemma~\ref{lem:surjective}:
We show that if a proper assignment set $A$ is given for a walk $W$ with $t(W)=1$, then a clockwise solution $S$ of $W$ with $\AS(S)=A$ exists.

If $A=\emptyset$ then since $A$ is a proper assignment set, this implies that $W$ has no PA-indices and $W$ itself is the only elementary cycle of $A,W$. 
This is a right-turn walk by Proposition~\ref{propo:elcyc_rightturn}, and since $t(W)=1$, it has a clockwise solution $S$ by Proposition~\ref{propo:turnnum_closedrightturnwalk}. Note that $\AS(S)=\emptyset=A$ since there are no PA-indices.

If $A\not=\emptyset$, then choose an arbitrary $\{x,y\}\in A$ with $n(y)=x$ (this exists since $A$ is a non-crossing perfect matching). We may again assume w.l.o.g. that $x=0$.
Use this to define $W'$, $W''$ and $A'$ as before.
$A'$ is a proper assignment set for $W'$ (Proposition~\ref{propo:newASisproper}).
Lemma~\ref{lem:inductionable} shows that for every \elcyc\ $C'$ of $W',A'$, an \elcyc\ $C$ of $W,A$ with $C\approx C'$ exists. By Lemma~\ref{lem:key_counting}, $t(C)=1$ follows from $t(W)=1$, so $t(C')=1$ for every such \elcyc\ $C'$. 
Therefore by applying Lemma~\ref{lem:key_counting} again, we obtain $t(W')=1$. 
At this point we have a new closed walk $W'$ in $\BW$ with turning number 1, with fewer PA-indices, and a proper assignment set $A'$ for it. Hence by induction, $W'$ admits a clockwise solution $S'$ with $\AS(S')=A'$.
$W''$ itself is an \elcyc\ of $W$, so $t(W'')=1$ (Lemma~\ref{lem:key_counting}), and therefore $W''$ admits a clockwise solution $S''$ as well (Proposition~\ref{propo:elcyc_rightturn}, Proposition~\ref{propo:turnnum_closedrightturnwalk}). 
Now Lemma~\ref{lem:gluing-lemma} shows that $S'$ and $S''$ can be combined into a clockwise solution $S$ for $W$ with $\AS(S)=A$.
\QED

\section{Proofs of Section~\ref{sec:circlegraphs}}

{\bf Proof} of Lemma~\ref{lem:Sroutine}:
We show that subroutine $S(i,j)$ correctly calculates $S_{i,j}$ when the stated $S_{x,y}$ values are known, in time $O(d(i))$.
Clearly, the algorithm only uses values $S_{x,y}$ with $-1\leq y-x<j-i$ for the calculations.

Observe that throughout the algorithm, the value of $m$ equals the size of some PM of $G_{i,j}$.
We now show that in some line, the size of a MPM is considered, which proves correctness.

Let $M$ be a MPM of $G_{i,j}$. If $j\leq i$ then $M=\emptyset$ which is considered in line~1. Now suppose $j>i$. If $M$ contains no edge incident with $i$, it is a PM of $G_{i+1,j}$ and considered in line~2. Otherwise, let $iv\in M$. If $v=j$, then $M$ consists of this edge and a PM of $G_{i+1,j-1}$, which is considered in line~6. Otherwise, because $M$ is non-crossing, it can be partitioned into a MPM of $G_{i,v}$ and $G_{v+1,j}$, which is considered in the for-loop.

The complexity of the algorithm is determined by the for-loop, which iterates at most $d(i)$ times.
\QED

{\bf Proof} of Lemma~\ref{lem:Nroutine}:
We show that subroutine $N(i,j)$ correctly calculates $N_{i,j}$ when the stated $N_{x,y}$ and $S_{x,y}$ values are known, in time $O(d(i))$.
The algorithm only uses values $S_{x,y}$ with $-1\leq y-x\leq j-i$ and values $N_{x,y}$ with $-1\leq y-x<j-i$ for the calculations. 

Clearly, line~1 returns the correct answer (in this case the empty set is the unique MPM). Otherwise, the algorithm adds $x$ to the number $N$ whenever $x$ different MPMs of $G_{i,j}$ are found that have not been considered earlier. We show that all cases are considered and no MPMs are double counted, which shows that the correct answer is returned in line~7.

If $S_{i,j}=S_{i+1,j}$ then all MPMs of $G_{i+1,j}$ are also MPMs of $G_{i,j}$, which explains line~3. This accounts for all MPMs of $G_{i,j}$ that do not contain an edge incident with $i$. $G_{i,j}$ admits MPMs $M$ that contain an edge $ij$ if and only if $S_{i,j}=w_{ij}+S_{i+1,j-1}$. Then $M-ij$ is a MPM of $G_{i+1,j-1}$, and there is a bijection between such MPMs, which explains line~6. It remains to consider MPMs $M$ of $G_{i,j}$ which contain an edge $iv$ with $i+1\leq v \leq j-1$. 
These can be decomposed into a PM of $G_{i,v}$ and a PM of $G_{v+1,j}$, which must be MPMs, so $S_{i,j}=S_{i,v}+S_{v+1,j}$. In fact, if $S_{i,j}=S_{i,v}+S_{v+1,j}$ then every combination of a MPM of $G_{i,v}$ and a MPM of $G_{v+1,j}$ gives a unique MPM of $G_{i,j}$ that contains $iv$, which explains line~5.
All cases are now considered, so at the end of the algorithm, $N=N_{i,j}$.

The complexity of the algorithm is determined by the for-loop, which iterates at most $d(i)$ times.
\QED

\end{document}

%% file: nontrivinstance.pstex_t
\begin{picture}(0,0)%
\includegraphics{nontrivinstance.pstex}%
\end{picture}%
\setlength{\unitlength}{3947sp}%
\begingroup\makeatletter\ifx\SetFigFontNFSS\undefined%
\gdef\SetFigFontNFSS#1#2#3#4#5{%
  \reset@font\fontsize{#1}{#2pt}%
  \fontfamily{#3}\fontseries{#4}\fontshape{#5}%
  \selectfont}%
\fi\endgroup%
\begin{picture}(6358,1451)(3572,-11565)
\put(3826,-10261){\makebox(0,0)[lb]{\smash{{\SetFigFontNFSS{10}{12.0}{\familydefault}{\mddefault}{\updefault}{\color[rgb]{0,0,0}$v_1$}%
}}}}
\put(4501,-10336){\makebox(0,0)[lb]{\smash{{\SetFigFontNFSS{10}{12.0}{\familydefault}{\mddefault}{\updefault}{\color[rgb]{0,0,0}$v_2$}%
}}}}
\put(7276,-10336){\makebox(0,0)[lb]{\smash{{\SetFigFontNFSS{10}{12.0}{\familydefault}{\mddefault}{\updefault}{\color[rgb]{0,0,0}$u_2$}%
}}}}
\put(7051,-10261){\makebox(0,0)[lb]{\smash{{\SetFigFontNFSS{10}{12.0}{\familydefault}{\mddefault}{\updefault}{\color[rgb]{0,0,0}$u_1$}%
}}}}
\end{picture}%

%% file: circlegraph.pstex_t
\begin{picture}(0,0)%
\includegraphics{circlegraph.pstex}%
\end{picture}%
\setlength{\unitlength}{3947sp}%
\begingroup\makeatletter\ifx\SetFigFontNFSS\undefined%
\gdef\SetFigFontNFSS#1#2#3#4#5{%
  \reset@font\fontsize{#1}{#2pt}%
  \fontfamily{#3}\fontseries{#4}\fontshape{#5}%
  \selectfont}%
\fi\endgroup%
\begin{picture}(4946,1540)(4096,-13074)
\put(8326,-11761){\makebox(0,0)[lb]{\smash{{\SetFigFontNFSS{8}{9.6}{\familydefault}{\mddefault}{\updefault}{\color[rgb]{0,0,0}1}%
}}}}
\put(8851,-11761){\makebox(0,0)[lb]{\smash{{\SetFigFontNFSS{8}{9.6}{\familydefault}{\mddefault}{\updefault}{\color[rgb]{0,0,0}4}%
}}}}
\put(8926,-12436){\makebox(0,0)[lb]{\smash{{\SetFigFontNFSS{8}{9.6}{\familydefault}{\mddefault}{\updefault}{\color[rgb]{0,0,0}6}%
}}}}
\put(7951,-12436){\makebox(0,0)[lb]{\smash{{\SetFigFontNFSS{8}{9.6}{\familydefault}{\mddefault}{\updefault}{\color[rgb]{0,0,0}3}%
}}}}
\put(4111,-12335){\makebox(0,0)[lb]{\smash{{\SetFigFontNFSS{8}{9.6}{\familydefault}{\mddefault}{\updefault}{\color[rgb]{0,0,0}$f$}%
}}}}
\put(5446,-11914){\makebox(0,0)[lb]{\smash{{\SetFigFontNFSS{8}{9.6}{\familydefault}{\mddefault}{\updefault}{\color[rgb]{0,0,0}$b$}%
}}}}
\put(5571,-12025){\makebox(0,0)[lb]{\smash{{\SetFigFontNFSS{8}{9.6}{\familydefault}{\mddefault}{\updefault}{\color[rgb]{0,0,0}$c$}%
}}}}
\put(5718,-11905){\makebox(0,0)[lb]{\smash{{\SetFigFontNFSS{8}{9.6}{\familydefault}{\mddefault}{\updefault}{\color[rgb]{0,0,0}$d$}%
}}}}
\put(5900,-12219){\makebox(0,0)[lb]{\smash{{\SetFigFontNFSS{8}{9.6}{\familydefault}{\mddefault}{\updefault}{\color[rgb]{0,0,0}$f$}%
}}}}
\put(4462,-12340){\makebox(0,0)[lb]{\smash{{\SetFigFontNFSS{8}{9.6}{\familydefault}{\mddefault}{\updefault}{\color[rgb]{0,0,0}$e$}%
}}}}
\put(4906,-12348){\makebox(0,0)[lb]{\smash{{\SetFigFontNFSS{8}{9.6}{\familydefault}{\mddefault}{\updefault}{\color[rgb]{0,0,0}$d$}%
}}}}
\put(4959,-11874){\makebox(0,0)[lb]{\smash{{\SetFigFontNFSS{8}{9.6}{\familydefault}{\mddefault}{\updefault}{\color[rgb]{0,0,0}$c$}%
}}}}
\put(4377,-11870){\makebox(0,0)[lb]{\smash{{\SetFigFontNFSS{8}{9.6}{\familydefault}{\mddefault}{\updefault}{\color[rgb]{0,0,0}$b$}%
}}}}
\put(4661,-11633){\makebox(0,0)[lb]{\smash{{\SetFigFontNFSS{8}{9.6}{\familydefault}{\mddefault}{\updefault}{\color[rgb]{0,0,0}$a$}%
}}}}
\put(8349,-12860){\makebox(0,0)[lb]{\smash{{\SetFigFontNFSS{8}{9.6}{\familydefault}{\mddefault}{\updefault}{\color[rgb]{0,0,0}model graph}%
}}}}
\put(7695,-12497){\makebox(0,0)[lb]{\smash{{\SetFigFontNFSS{8}{9.6}{\familydefault}{\mddefault}{\updefault}{\color[rgb]{0,0,0}4}%
}}}}
\put(7321,-12431){\makebox(0,0)[lb]{\smash{{\SetFigFontNFSS{8}{9.6}{\familydefault}{\mddefault}{\updefault}{\color[rgb]{0,0,0}5}%
}}}}
\put(7929,-11774){\makebox(0,0)[lb]{\smash{{\SetFigFontNFSS{8}{9.6}{\familydefault}{\mddefault}{\updefault}{\color[rgb]{0,0,0}2}%
}}}}
\put(7696,-11735){\makebox(0,0)[lb]{\smash{{\SetFigFontNFSS{8}{9.6}{\familydefault}{\mddefault}{\updefault}{\color[rgb]{0,0,0}1}%
}}}}
\put(7325,-11778){\makebox(0,0)[lb]{\smash{{\SetFigFontNFSS{8}{9.6}{\familydefault}{\mddefault}{\updefault}{\color[rgb]{0,0,0}0}%
}}}}
\put(8194,-11911){\makebox(0,0)[lb]{\smash{{\SetFigFontNFSS{8}{9.6}{\familydefault}{\mddefault}{\updefault}{\color[rgb]{0,0,0}0}%
}}}}
\put(8551,-11660){\makebox(0,0)[lb]{\smash{{\SetFigFontNFSS{8}{9.6}{\familydefault}{\mddefault}{\updefault}{\color[rgb]{0,0,0}2}%
}}}}
\put(8705,-11660){\makebox(0,0)[lb]{\smash{{\SetFigFontNFSS{8}{9.6}{\familydefault}{\mddefault}{\updefault}{\color[rgb]{0,0,0}3}%
}}}}
\put(8966,-11885){\makebox(0,0)[lb]{\smash{{\SetFigFontNFSS{8}{9.6}{\familydefault}{\mddefault}{\updefault}{\color[rgb]{0,0,0}5}%
}}}}
\put(8736,-12493){\makebox(0,0)[lb]{\smash{{\SetFigFontNFSS{8}{9.6}{\familydefault}{\mddefault}{\updefault}{\color[rgb]{0,0,0}7}%
}}}}
\put(8467,-12507){\makebox(0,0)[lb]{\smash{{\SetFigFontNFSS{8}{9.6}{\familydefault}{\mddefault}{\updefault}{\color[rgb]{0,0,0}9}%
}}}}
\put(8599,-12560){\makebox(0,0)[lb]{\smash{{\SetFigFontNFSS{8}{9.6}{\familydefault}{\mddefault}{\updefault}{\color[rgb]{0,0,0}8}%
}}}}
\put(8255,-12449){\makebox(0,0)[lb]{\smash{{\SetFigFontNFSS{8}{9.6}{\familydefault}{\mddefault}{\updefault}{\color[rgb]{0,0,0}10}%
}}}}
\put(8145,-12260){\makebox(0,0)[lb]{\smash{{\SetFigFontNFSS{8}{9.6}{\familydefault}{\mddefault}{\updefault}{\color[rgb]{0,0,0}11}%
}}}}
\put(5795,-12080){\makebox(0,0)[lb]{\smash{{\SetFigFontNFSS{8}{9.6}{\familydefault}{\mddefault}{\updefault}{\color[rgb]{0,0,0}$e$}%
}}}}
\put(5346,-12090){\makebox(0,0)[lb]{\smash{{\SetFigFontNFSS{8}{9.6}{\familydefault}{\mddefault}{\updefault}{\color[rgb]{0,0,0}$a$}%
}}}}
\put(8295,-12703){\makebox(0,0)[lb]{\smash{{\SetFigFontNFSS{8}{9.6}{\familydefault}{\mddefault}{\updefault}{\color[rgb]{0,0,0}simple chord}%
}}}}
\put(7526,-12859){\makebox(0,0)[lb]{\smash{{\SetFigFontNFSS{8}{9.6}{\familydefault}{\mddefault}{\updefault}{\color[rgb]{0,0,0}graph}%
}}}}
\put(7320,-12708){\makebox(0,0)[lb]{\smash{{\SetFigFontNFSS{8}{9.6}{\familydefault}{\mddefault}{\updefault}{\color[rgb]{0,0,0}chord model}%
}}}}
\put(6395,-12870){\makebox(0,0)[lb]{\smash{{\SetFigFontNFSS{8}{9.6}{\familydefault}{\mddefault}{\updefault}{\color[rgb]{0,0,0}diagram}%
}}}}
\put(6251,-12704){\makebox(0,0)[lb]{\smash{{\SetFigFontNFSS{8}{9.6}{\familydefault}{\mddefault}{\updefault}{\color[rgb]{0,0,0}simple chord}%
}}}}
\put(4343,-12702){\makebox(0,0)[lb]{\smash{{\SetFigFontNFSS{8}{9.6}{\familydefault}{\mddefault}{\updefault}{\color[rgb]{0,0,0}circle graph}%
}}}}
\put(5279,-12709){\makebox(0,0)[lb]{\smash{{\SetFigFontNFSS{8}{9.6}{\familydefault}{\mddefault}{\updefault}{\color[rgb]{0,0,0}chord diagram}%
}}}}
\put(6247,-12090){\makebox(0,0)[lb]{\smash{{\SetFigFontNFSS{8}{9.6}{\familydefault}{\mddefault}{\updefault}{\color[rgb]{0,0,0}$a$}%
}}}}
\put(6405,-12260){\makebox(0,0)[lb]{\smash{{\SetFigFontNFSS{8}{9.6}{\familydefault}{\mddefault}{\updefault}{\color[rgb]{0,0,0}$b$}%
}}}}
\put(6428,-11896){\makebox(0,0)[lb]{\smash{{\SetFigFontNFSS{8}{9.6}{\familydefault}{\mddefault}{\updefault}{\color[rgb]{0,0,0}$c$}%
}}}}
\put(6615,-11972){\makebox(0,0)[lb]{\smash{{\SetFigFontNFSS{8}{9.6}{\familydefault}{\mddefault}{\updefault}{\color[rgb]{0,0,0}$d$}%
}}}}
\put(6683,-12258){\makebox(0,0)[lb]{\smash{{\SetFigFontNFSS{8}{9.6}{\familydefault}{\mddefault}{\updefault}{\color[rgb]{0,0,0}$e$}%
}}}}
\put(6881,-12108){\makebox(0,0)[lb]{\smash{{\SetFigFontNFSS{8}{9.6}{\familydefault}{\mddefault}{\updefault}{\color[rgb]{0,0,0}$f$}%
}}}}
\put(8584,-13014){\makebox(0,0)[lb]{\smash{{\SetFigFontNFSS{8}{9.6}{\familydefault}{\mddefault}{\updefault}{\color[rgb]{0,0,0}(e)}%
}}}}
\put(7585,-13014){\makebox(0,0)[lb]{\smash{{\SetFigFontNFSS{8}{9.6}{\familydefault}{\mddefault}{\updefault}{\color[rgb]{0,0,0}(d)}%
}}}}
\put(6531,-13015){\makebox(0,0)[lb]{\smash{{\SetFigFontNFSS{8}{9.6}{\familydefault}{\mddefault}{\updefault}{\color[rgb]{0,0,0}(c)}%
}}}}
\put(5631,-13028){\makebox(0,0)[lb]{\smash{{\SetFigFontNFSS{8}{9.6}{\familydefault}{\mddefault}{\updefault}{\color[rgb]{0,0,0}(b)}%
}}}}
\put(4605,-13020){\makebox(0,0)[lb]{\smash{{\SetFigFontNFSS{8}{9.6}{\familydefault}{\mddefault}{\updefault}{\color[rgb]{0,0,0}(a)}%
}}}}
\end{picture}%

%% file: brickwall_nolabels.pstex_t
\begin{picture}(0,0)%
\includegraphics{brickwall_nolabels.pstex}%
\end{picture}%
\setlength{\unitlength}{3947sp}%
\begingroup\makeatletter\ifx\SetFigFontNFSS\undefined%
\gdef\SetFigFontNFSS#1#2#3#4#5{%
  \reset@font\fontsize{#1}{#2pt}%
  \fontfamily{#3}\fontseries{#4}\fontshape{#5}%
  \selectfont}%
\fi\endgroup%
\begin{picture}(5424,1824)(3439,-12748)
\end{picture}%

%% file: PA-indB.pstex_t
\begin{picture}(0,0)%
\includegraphics{PA-indB.pstex}%
\end{picture}%
\setlength{\unitlength}{3947sp}%
\begingroup\makeatletter\ifx\SetFigFontNFSS\undefined%
\gdef\SetFigFontNFSS#1#2#3#4#5{%
  \reset@font\fontsize{#1}{#2pt}%
  \fontfamily{#3}\fontseries{#4}\fontshape{#5}%
  \selectfont}%
\fi\endgroup%
\begin{picture}(6860,2476)(278,-13400)
\put(526,-12511){\makebox(0,0)[lb]{\smash{{\SetFigFontNFSS{10}{12.0}{\familydefault}{\mddefault}{\updefault}{\color[rgb]{0,0,0}0}%
}}}}
\put(2371,-11501){\makebox(0,0)[lb]{\smash{{\SetFigFontNFSS{10}{12.0}{\familydefault}{\mddefault}{\updefault}{\color[rgb]{0,0,0}13}%
}}}}
\put(2486,-11700){\makebox(0,0)[lb]{\smash{{\SetFigFontNFSS{10}{12.0}{\familydefault}{\mddefault}{\updefault}{\color[rgb]{0,0,0}14}%
}}}}
\put(2069,-12615){\makebox(0,0)[lb]{\smash{{\SetFigFontNFSS{10}{12.0}{\familydefault}{\mddefault}{\updefault}{\color[rgb]{0,0,0}21}%
}}}}
\put(1706,-12054){\makebox(0,0)[lb]{\smash{{\SetFigFontNFSS{10}{12.0}{\familydefault}{\mddefault}{\updefault}{\color[rgb]{0,0,0}24}%
}}}}
\put(1551,-11889){\makebox(0,0)[lb]{\smash{{\SetFigFontNFSS{10}{12.0}{\familydefault}{\mddefault}{\updefault}{\color[rgb]{0,0,0}25}%
}}}}
\put(1137,-11847){\makebox(0,0)[lb]{\smash{{\SetFigFontNFSS{10}{12.0}{\familydefault}{\mddefault}{\updefault}{\color[rgb]{0,0,0}27}%
}}}}
\put(1066,-12030){\makebox(0,0)[lb]{\smash{{\SetFigFontNFSS{10}{12.0}{\familydefault}{\mddefault}{\updefault}{\color[rgb]{0,0,0}28}%
}}}}
\put(1501,-12586){\makebox(0,0)[lb]{\smash{{\SetFigFontNFSS{10}{12.0}{\familydefault}{\mddefault}{\updefault}{\color[rgb]{0,0,0}32}%
}}}}
\put(4031,-11534){\makebox(0,0)[lb]{\smash{{\SetFigFontNFSS{10}{12.0}{\familydefault}{\mddefault}{\updefault}{\color[rgb]{0,0,0}4}%
}}}}
\put(5611,-11535){\makebox(0,0)[lb]{\smash{{\SetFigFontNFSS{10}{12.0}{\familydefault}{\mddefault}{\updefault}{\color[rgb]{0,0,0}25}%
}}}}
\put(4510,-11521){\makebox(0,0)[lb]{\smash{{\SetFigFontNFSS{10}{12.0}{\familydefault}{\mddefault}{\updefault}{\color[rgb]{0,0,0}27}%
}}}}
\put(6240,-11986){\makebox(0,0)[lb]{\smash{{\SetFigFontNFSS{10}{12.0}{\familydefault}{\mddefault}{\updefault}{\color[rgb]{0,0,0}14}%
}}}}
\put(6331,-11521){\makebox(0,0)[lb]{\smash{{\SetFigFontNFSS{10}{12.0}{\familydefault}{\mddefault}{\updefault}{\color[rgb]{0,0,0}13}%
}}}}
\put(5626,-11987){\makebox(0,0)[lb]{\smash{{\SetFigFontNFSS{10}{12.0}{\familydefault}{\mddefault}{\updefault}{\color[rgb]{0,0,0}24}%
}}}}
\put(5596,-12415){\makebox(0,0)[lb]{\smash{{\SetFigFontNFSS{10}{12.0}{\familydefault}{\mddefault}{\updefault}{\color[rgb]{0,0,0}21}%
}}}}
\put(6239,-12423){\makebox(0,0)[lb]{\smash{{\SetFigFontNFSS{10}{12.0}{\familydefault}{\mddefault}{\updefault}{\color[rgb]{0,0,0}17}%
}}}}
\put(5924,-12728){\makebox(0,0)[lb]{\smash{{\SetFigFontNFSS{10}{12.0}{\familydefault}{\mddefault}{\updefault}{\color[rgb]{0,0,0}32}%
}}}}
\put(4183,-12754){\makebox(0,0)[lb]{\smash{{\SetFigFontNFSS{10}{12.0}{\familydefault}{\mddefault}{\updefault}{\color[rgb]{0,0,0}0}%
}}}}
\put(4566,-12297){\makebox(0,0)[lb]{\smash{{\SetFigFontNFSS{10}{12.0}{\familydefault}{\mddefault}{\updefault}{\color[rgb]{0,0,0}28}%
}}}}
\put(4247,-12291){\makebox(0,0)[lb]{\smash{{\SetFigFontNFSS{10}{12.0}{\familydefault}{\mddefault}{\updefault}{\color[rgb]{0,0,0}1}%
}}}}
\put(541,-12125){\makebox(0,0)[lb]{\smash{{\SetFigFontNFSS{10}{12.0}{\familydefault}{\mddefault}{\updefault}{\color[rgb]{0,0,0}1}%
}}}}
\put(657,-11494){\makebox(0,0)[lb]{\smash{{\SetFigFontNFSS{10}{12.0}{\familydefault}{\mddefault}{\updefault}{\color[rgb]{0,0,0}4}%
}}}}
\put(2805,-12269){\makebox(0,0)[lb]{\smash{{\SetFigFontNFSS{10}{12.0}{\familydefault}{\mddefault}{\updefault}{\color[rgb]{0,0,0}17}%
}}}}
\put(3151,-11611){\makebox(0,0)[lb]{\smash{{\SetFigFontNFSS{10}{12.0}{\familydefault}{\mddefault}{\updefault}{\color[rgb]{0,0,0}$\phi$}%
}}}}
\put(526,-13336){\makebox(0,0)[lb]{\smash{{\SetFigFontNFSS{10}{12.0}{\familydefault}{\mddefault}{\updefault}{\color[rgb]{0,0,0}Assignment paths in the patch $H$ }%
}}}}
\put(4651,-13336){\makebox(0,0)[lb]{\smash{{\SetFigFontNFSS{10}{12.0}{\familydefault}{\mddefault}{\updefault}{\color[rgb]{0,0,0}PA-indices of the walk $W$}%
}}}}
\put(293,-13324){\makebox(0,0)[lb]{\smash{{\SetFigFontNFSS{10}{12.0}{\familydefault}{\mddefault}{\updefault}{\color[rgb]{0,0,0}(a)}%
}}}}
\put(4390,-13333){\makebox(0,0)[lb]{\smash{{\SetFigFontNFSS{10}{12.0}{\familydefault}{\mddefault}{\updefault}{\color[rgb]{0,0,0}(b)}%
}}}}
\end{picture}%

%% file: elcycsB.pstex_t
\begin{picture}(0,0)%
\includegraphics{elcycsB.pstex}%
\end{picture}%
\setlength{\unitlength}{3947sp}%
\begingroup\makeatletter\ifx\SetFigFontNFSS\undefined%
\gdef\SetFigFontNFSS#1#2#3#4#5{%
  \reset@font\fontsize{#1}{#2pt}%
  \fontfamily{#3}\fontseries{#4}\fontshape{#5}%
  \selectfont}%
\fi\endgroup%
\begin{picture}(7166,2469)(3514,-13393)
\put(4268,-12285){\makebox(0,0)[lb]{\smash{{\SetFigFontNFSS{10}{12.0}{\familydefault}{\mddefault}{\updefault}{\color[rgb]{0,0,0}1}%
}}}}
\put(4166,-12749){\makebox(0,0)[lb]{\smash{{\SetFigFontNFSS{10}{12.0}{\familydefault}{\mddefault}{\updefault}{\color[rgb]{0,0,0}0}%
}}}}
\put(4546,-12283){\makebox(0,0)[lb]{\smash{{\SetFigFontNFSS{10}{12.0}{\familydefault}{\mddefault}{\updefault}{\color[rgb]{0,0,0}28}%
}}}}
\put(4564,-11501){\makebox(0,0)[lb]{\smash{{\SetFigFontNFSS{10}{12.0}{\familydefault}{\mddefault}{\updefault}{\color[rgb]{0,0,0}27}%
}}}}
\put(4071,-11521){\makebox(0,0)[lb]{\smash{{\SetFigFontNFSS{10}{12.0}{\familydefault}{\mddefault}{\updefault}{\color[rgb]{0,0,0}4}%
}}}}
\put(5471,-11508){\makebox(0,0)[lb]{\smash{{\SetFigFontNFSS{10}{12.0}{\familydefault}{\mddefault}{\updefault}{\color[rgb]{0,0,0}25}%
}}}}
\put(6371,-11501){\makebox(0,0)[lb]{\smash{{\SetFigFontNFSS{10}{12.0}{\familydefault}{\mddefault}{\updefault}{\color[rgb]{0,0,0}13}%
}}}}
\put(6239,-11980){\makebox(0,0)[lb]{\smash{{\SetFigFontNFSS{10}{12.0}{\familydefault}{\mddefault}{\updefault}{\color[rgb]{0,0,0}14}%
}}}}
\put(5626,-11980){\makebox(0,0)[lb]{\smash{{\SetFigFontNFSS{10}{12.0}{\familydefault}{\mddefault}{\updefault}{\color[rgb]{0,0,0}24}%
}}}}
\put(5422,-12408){\makebox(0,0)[lb]{\smash{{\SetFigFontNFSS{10}{12.0}{\familydefault}{\mddefault}{\updefault}{\color[rgb]{0,0,0}21}%
}}}}
\put(6432,-12403){\makebox(0,0)[lb]{\smash{{\SetFigFontNFSS{10}{12.0}{\familydefault}{\mddefault}{\updefault}{\color[rgb]{0,0,0}17}%
}}}}
\put(5937,-12748){\makebox(0,0)[lb]{\smash{{\SetFigFontNFSS{10}{12.0}{\familydefault}{\mddefault}{\updefault}{\color[rgb]{0,0,0}32}%
}}}}
\put(4337,-13329){\makebox(0,0)[lb]{\smash{{\SetFigFontNFSS{10}{12.0}{\familydefault}{\mddefault}{\updefault}{\color[rgb]{0,0,0}Walk $W$  and proper assignment $A$}%
}}}}
\put(7676,-13320){\makebox(0,0)[lb]{\smash{{\SetFigFontNFSS{10}{12.0}{\familydefault}{\mddefault}{\updefault}{\color[rgb]{0,0,0}(b)}%
}}}}
\put(4126,-13318){\makebox(0,0)[lb]{\smash{{\SetFigFontNFSS{10}{12.0}{\familydefault}{\mddefault}{\updefault}{\color[rgb]{0,0,0}(a)}%
}}}}
\put(7883,-13315){\makebox(0,0)[lb]{\smash{{\SetFigFontNFSS{10}{12.0}{\familydefault}{\mddefault}{\updefault}{\color[rgb]{0,0,0}The elementary cycles of $W,A$}%
}}}}
\end{picture}%

%% file: rightturn.pstex_t
\begin{picture}(0,0)%
\includegraphics{rightturn.pstex}%
\end{picture}%
\setlength{\unitlength}{3947sp}%
\begingroup\makeatletter\ifx\SetFigFontNFSS\undefined%
\gdef\SetFigFontNFSS#1#2#3#4#5{%
  \reset@font\fontsize{#1}{#2pt}%
  \fontfamily{#3}\fontseries{#4}\fontshape{#5}%
  \selectfont}%
\fi\endgroup%
\begin{picture}(3252,984)(3886,-12748)
\put(4576,-12436){\makebox(0,0)[lb]{\smash{{\SetFigFontNFSS{10}{12.0}{\familydefault}{\mddefault}{\updefault}{\color[rgb]{0,0,0}$b_{2,0}$}%
}}}}
\put(5476,-12211){\makebox(0,0)[lb]{\smash{{\SetFigFontNFSS{10}{12.0}{\familydefault}{\mddefault}{\updefault}{\color[rgb]{0,0,0}$b_{5,1}$}%
}}}}
\put(5476,-12361){\makebox(0,0)[lb]{\smash{{\SetFigFontNFSS{10}{12.0}{\familydefault}{\mddefault}{\updefault}{\color[rgb]{0,0,0}$b_{4,0}$}%
}}}}
\put(6376,-12211){\makebox(0,0)[lb]{\smash{{\SetFigFontNFSS{10}{12.0}{\familydefault}{\mddefault}{\updefault}{\color[rgb]{0,0,0}$b_{7,1}$}%
}}}}
\put(6376,-12436){\makebox(0,0)[lb]{\smash{{\SetFigFontNFSS{10}{12.0}{\familydefault}{\mddefault}{\updefault}{\color[rgb]{0,0,0}$b_{6,0}$}%
}}}}
\put(4126,-11986){\makebox(0,0)[lb]{\smash{{\SetFigFontNFSS{10}{12.0}{\familydefault}{\mddefault}{\updefault}{\color[rgb]{0,0,0}$b_{2,1}$}%
}}}}
\put(4126,-12661){\makebox(0,0)[lb]{\smash{{\SetFigFontNFSS{10}{12.0}{\familydefault}{\mddefault}{\updefault}{\color[rgb]{0,0,0}$b_{1,0}$}%
}}}}
\put(4576,-12211){\makebox(0,0)[lb]{\smash{{\SetFigFontNFSS{10}{12.0}{\familydefault}{\mddefault}{\updefault}{\color[rgb]{0,0,0}$b_{3,1}$}%
}}}}
\put(3901,-12436){\makebox(0,0)[lb]{\smash{{\SetFigFontNFSS{10}{12.0}{\familydefault}{\mddefault}{\updefault}{\color[rgb]{0,0,0}$W_0$}%
}}}}
\put(3901,-12211){\makebox(0,0)[lb]{\smash{{\SetFigFontNFSS{10}{12.0}{\familydefault}{\mddefault}{\updefault}{\color[rgb]{0,0,0}$W_1$}%
}}}}
\put(5926,-12661){\makebox(0,0)[lb]{\smash{{\SetFigFontNFSS{10}{12.0}{\familydefault}{\mddefault}{\updefault}{\color[rgb]{0,0,0}$b_{5,0}$}%
}}}}
\put(5026,-12661){\makebox(0,0)[lb]{\smash{{\SetFigFontNFSS{10}{12.0}{\familydefault}{\mddefault}{\updefault}{\color[rgb]{0,0,0}$b_{3,0}$}%
}}}}
\put(5026,-11911){\makebox(0,0)[lb]{\smash{{\SetFigFontNFSS{10}{12.0}{\familydefault}{\mddefault}{\updefault}{\color[rgb]{0,0,0}$b_{4,1}$}%
}}}}
\put(5926,-11911){\makebox(0,0)[lb]{\smash{{\SetFigFontNFSS{10}{12.0}{\familydefault}{\mddefault}{\updefault}{\color[rgb]{0,0,0}$b_{6,1}$}%
}}}}
\put(6826,-12661){\makebox(0,0)[lb]{\smash{{\SetFigFontNFSS{10}{12.0}{\familydefault}{\mddefault}{\updefault}{\color[rgb]{0,0,0}$b_{7,0}$}%
}}}}
\put(6826,-11911){\makebox(0,0)[lb]{\smash{{\SetFigFontNFSS{10}{12.0}{\familydefault}{\mddefault}{\updefault}{\color[rgb]{0,0,0}$b_{8,1}$}%
}}}}
\end{picture}%

%% file: splitB.pstex_t
\begin{picture}(0,0)%
\includegraphics{splitB.pstex}%
\end{picture}%
\setlength{\unitlength}{3947sp}%
\begingroup\makeatletter\ifx\SetFigFontNFSS\undefined%
\gdef\SetFigFontNFSS#1#2#3#4#5{%
  \reset@font\fontsize{#1}{#2pt}%
  \fontfamily{#3}\fontseries{#4}\fontshape{#5}%
  \selectfont}%
\fi\endgroup%
\begin{picture}(6999,2692)(3514,-13616)
\put(4183,-12754){\makebox(0,0)[lb]{\smash{{\SetFigFontNFSS{10}{12.0}{\familydefault}{\mddefault}{\updefault}{\color[rgb]{0,0,0}0}%
}}}}
\put(5436,-12408){\makebox(0,0)[lb]{\smash{{\SetFigFontNFSS{10}{12.0}{\familydefault}{\mddefault}{\updefault}{\color[rgb]{0,0,0}21}%
}}}}
\put(5811,-12728){\makebox(0,0)[lb]{\smash{{\SetFigFontNFSS{10}{12.0}{\familydefault}{\mddefault}{\updefault}{\color[rgb]{0,0,0}32=$y$}%
}}}}
\put(4031,-11534){\makebox(0,0)[lb]{\smash{{\SetFigFontNFSS{10}{12.0}{\familydefault}{\mddefault}{\updefault}{\color[rgb]{0,0,0}4}%
}}}}
\put(5611,-11535){\makebox(0,0)[lb]{\smash{{\SetFigFontNFSS{10}{12.0}{\familydefault}{\mddefault}{\updefault}{\color[rgb]{0,0,0}25}%
}}}}
\put(4510,-11521){\makebox(0,0)[lb]{\smash{{\SetFigFontNFSS{10}{12.0}{\familydefault}{\mddefault}{\updefault}{\color[rgb]{0,0,0}27}%
}}}}
\put(6240,-11986){\makebox(0,0)[lb]{\smash{{\SetFigFontNFSS{10}{12.0}{\familydefault}{\mddefault}{\updefault}{\color[rgb]{0,0,0}14}%
}}}}
\put(6331,-11521){\makebox(0,0)[lb]{\smash{{\SetFigFontNFSS{10}{12.0}{\familydefault}{\mddefault}{\updefault}{\color[rgb]{0,0,0}13}%
}}}}
\put(5626,-11987){\makebox(0,0)[lb]{\smash{{\SetFigFontNFSS{10}{12.0}{\familydefault}{\mddefault}{\updefault}{\color[rgb]{0,0,0}24}%
}}}}
\put(4566,-12297){\makebox(0,0)[lb]{\smash{{\SetFigFontNFSS{10}{12.0}{\familydefault}{\mddefault}{\updefault}{\color[rgb]{0,0,0}28}%
}}}}
\put(4247,-12291){\makebox(0,0)[lb]{\smash{{\SetFigFontNFSS{10}{12.0}{\familydefault}{\mddefault}{\updefault}{\color[rgb]{0,0,0}1}%
}}}}
\put(6412,-12403){\makebox(0,0)[lb]{\smash{{\SetFigFontNFSS{10}{12.0}{\familydefault}{\mddefault}{\updefault}{\color[rgb]{0,0,0}17}%
}}}}
\put(9751,-12736){\makebox(0,0)[lb]{\smash{{\SetFigFontNFSS{10}{12.0}{\familydefault}{\mddefault}{\updefault}{\color[rgb]{0,0,0}0}%
}}}}
\put(8008,-12754){\makebox(0,0)[lb]{\smash{{\SetFigFontNFSS{10}{12.0}{\familydefault}{\mddefault}{\updefault}{\color[rgb]{0,0,0}10}%
}}}}
\put(4733,-13321){\makebox(0,0)[lb]{\smash{{\SetFigFontNFSS{10}{12.0}{\familydefault}{\mddefault}{\updefault}{\color[rgb]{0,0,0}$W'=W_{0,32}\circ H_{32,0}$}%
}}}}
\put(5101,-13561){\makebox(0,0)[lb]{\smash{{\SetFigFontNFSS{10}{12.0}{\familydefault}{\mddefault}{\updefault}{\color[rgb]{0,0,0}: $A'$}%
}}}}
\put(7726,-13336){\makebox(0,0)[lb]{\smash{{\SetFigFontNFSS{10}{12.0}{\familydefault}{\mddefault}{\updefault}{\color[rgb]{0,0,0}$W''=W_{32,0}\circ H_{0,23}$}%
}}}}
\put(10126,-12736){\makebox(0,0)[lb]{\smash{{\SetFigFontNFSS{10}{12.0}{\familydefault}{\mddefault}{\updefault}{\color[rgb]{0,0,0}1}%
}}}}
\end{picture}%

%% file: glueB.pstex_t
\begin{picture}(0,0)%
\includegraphics{glueB.pstex}%
\end{picture}%
\setlength{\unitlength}{3947sp}%
\begingroup\makeatletter\ifx\SetFigFontNFSS\undefined%
\gdef\SetFigFontNFSS#1#2#3#4#5{%
  \reset@font\fontsize{#1}{#2pt}%
  \fontfamily{#3}\fontseries{#4}\fontshape{#5}%
  \selectfont}%
\fi\endgroup%
\begin{picture}(6508,2608)(-235,-13590)
\put(1351,-12661){\makebox(0,0)[lb]{\smash{{\SetFigFontNFSS{8}{9.6}{\familydefault}{\mddefault}{\updefault}{\color[rgb]{0,0,0}$v'_1$}%
}}}}
\put(835,-12661){\makebox(0,0)[lb]{\smash{{\SetFigFontNFSS{8}{9.6}{\familydefault}{\mddefault}{\updefault}{\color[rgb]{0,0,0}$v'_3$}%
}}}}
\put(1054,-12557){\makebox(0,0)[lb]{\smash{{\SetFigFontNFSS{8}{9.6}{\familydefault}{\mddefault}{\updefault}{\color[rgb]{0,0,0}$v'_2$}%
}}}}
\put(993,-12943){\makebox(0,0)[lb]{\smash{{\SetFigFontNFSS{8}{9.6}{\familydefault}{\mddefault}{\updefault}{\color[rgb]{0,0,0}$v''_2$}%
}}}}
\put(143,-12884){\makebox(0,0)[lb]{\smash{{\SetFigFontNFSS{8}{9.6}{\familydefault}{\mddefault}{\updefault}{\color[rgb]{0,0,0}$\Bt_x=v''_0$}%
}}}}
\put(4951,-12736){\makebox(0,0)[lb]{\smash{{\SetFigFontNFSS{8}{9.6}{\familydefault}{\mddefault}{\updefault}{\color[rgb]{0,0,0}$v_0$}%
}}}}
\put(4378,-12925){\makebox(0,0)[lb]{\smash{{\SetFigFontNFSS{8}{9.6}{\familydefault}{\mddefault}{\updefault}{\color[rgb]{0,0,0}$v_2$}%
}}}}
\put(3811,-12610){\makebox(0,0)[lb]{\smash{{\SetFigFontNFSS{8}{9.6}{\familydefault}{\mddefault}{\updefault}{\color[rgb]{0,0,0}$v_4$}%
}}}}
\put(788,-13038){\makebox(0,0)[lb]{\smash{{\SetFigFontNFSS{8}{9.6}{\familydefault}{\mddefault}{\updefault}{\color[rgb]{0,0,0}$v''_1$}%
}}}}
\put(1220,-13051){\makebox(0,0)[lb]{\smash{{\SetFigFontNFSS{8}{9.6}{\familydefault}{\mddefault}{\updefault}{\color[rgb]{0,0,0}$v''_3$}%
}}}}
\put(176,-12431){\makebox(0,0)[lb]{\smash{{\SetFigFontNFSS{8}{9.6}{\familydefault}{\mddefault}{\updefault}{\color[rgb]{0,0,0}$\Bo_0=v'_4$}%
}}}}
\put(4011,-12934){\makebox(0,0)[lb]{\smash{{\SetFigFontNFSS{8}{9.6}{\familydefault}{\mddefault}{\updefault}{\color[rgb]{0,0,0}$v_3$}%
}}}}
\put(4723,-12964){\makebox(0,0)[lb]{\smash{{\SetFigFontNFSS{8}{9.6}{\familydefault}{\mddefault}{\updefault}{\color[rgb]{0,0,0}$v_1$}%
}}}}
\put(1629,-12384){\makebox(0,0)[lb]{\smash{{\SetFigFontNFSS{8}{9.6}{\familydefault}{\mddefault}{\updefault}{\color[rgb]{0,0,0}$\Bo_y=$}%
}}}}
\put(1487,-12938){\makebox(0,0)[lb]{\smash{{\SetFigFontNFSS{8}{9.6}{\familydefault}{\mddefault}{\updefault}{\color[rgb]{0,0,0}$v''_4=\Bt_0$}%
}}}}
\put(1622,-12561){\makebox(0,0)[lb]{\smash{{\SetFigFontNFSS{8}{9.6}{\familydefault}{\mddefault}{\updefault}{\color[rgb]{0,0,0}$v'_0$}%
}}}}
\put(-220,-13259){\makebox(0,0)[lb]{\smash{{\SetFigFontNFSS{10}{12.0}{\familydefault}{\mddefault}{\updefault}{\color[rgb]{0,0,0}$H''$:}%
}}}}
\put(-219,-11786){\makebox(0,0)[lb]{\smash{{\SetFigFontNFSS{10}{12.0}{\familydefault}{\mddefault}{\updefault}{\color[rgb]{0,0,0}$H'$:}%
}}}}
\put(3414,-11793){\makebox(0,0)[lb]{\smash{{\SetFigFontNFSS{10}{12.0}{\familydefault}{\mddefault}{\updefault}{\color[rgb]{0,0,0}$H$:}%
}}}}
\end{picture}%